\documentclass[prd,amsmath,amssymb,letterpaper]{revtex4}
\usepackage{graphicx}    
\usepackage{bm}
\usepackage{epsf}
\usepackage{color}

\begin{document}

\title{Confronting the short-baseline oscillation anomalies with a single sterile neutrino and non-standard matter effects}

\author{G.~Karagiorgi$^1$}
\email{georgia@nevis.columbia.edu}
\author{M.~H.~Shaevitz$^1$}
\email{shaevitz@nevis.columbia.edu}
\author{J.~M.~Conrad$^2$}
\email{conrad@mit.edu}

\affiliation{$^1$Department of Physics, Columbia University, New York, NY 10027}
\affiliation{$^2$Massachusetts Institute of Technology, Cambridge, MA 02139}

\smallskip

\date{\today}

\begin{abstract}
We examine the MiniBooNE neutrino, MiniBooNE antineutrino and LSND antineutrino data sets in a two-neutrino $\stackrel{\tiny{(-)}}{\nu}_{\mu}\rightarrow\stackrel{\tiny{(-)}}{\nu}_e$ oscillation approximation subject to non-standard matter effects. We assume those effects can be parametrized by an $L$-independent effective potential, $V_s=\pm A_s$, experienced only by an intermediate, non-weakly-interacting (sterile) neutrino state which we assume participates in the oscillation, where $+/-$ corresponds to neutrino/antineutrino propagation. We discuss the mathematical framework in which such oscillations arise in detail, and derive the relevant oscillation probability as a function of the vacuum oscillation parameters $\Delta m^2$ and $\sin^22\theta_{\mu e}$, and the matter effect parameter $A_s$. We are able to successfully fit all three data sets, including the MiniBooNE low energy excess, with the following best-fit model parameters: $\Delta m^2=0.47$~eV$^2$, $\sin^22\theta_{\mu e}=0.010$, and $A_s=2.0\times10^{-10}$~eV. The $\chi^2$-probability for the best fit corresponds to 21.6\%, to be compared to 6.8\% for a fit where $A_s$ has been set to zero, corresponding to a (3+1) sterile neutrino oscillation model. We find that the compatibility between the three data sets corresponds to 17.4\%, to be compared to 2.3\% for $A_s=0$. 
Finally, given the fit results, we examine consequences for reactor, solar, and atmospheric oscillations. For this paper, the presented model is empirically driven, but the results obtained can be directly used to investigate various phenomenological interpretations such as non-standard matter effects.
\end{abstract}


\maketitle

\section{\label{sec:one}INTRODUCTION}

Neutrino flavor oscillation is a phenomenon that arises due to non-zero, distinct neutrino masses and leptonic mixing. The current Standard Model picture incorporating neutrino oscillation relates three neutrino mass eigenstates with masses $m_1$, $m_2$ and $m_3$ to the three neutrino weak flavor eigenstates, $\nu_e$, $\nu_{\mu}$, and $\nu_{\tau}$. This is done through the leptonic mixing matrix, $U$, defined as a product of three rotations,
\begin{eqnarray}
\left(
\begin{array}{c}
\nu_e \\
\nu_\mu \\
\nu_\tau
\end{array} 
\right)
&=&
\left(
\begin{array}{ccc}
U_{e1} & U_{e2} & U_{e3} \\
U_{\mu1} & U_{\mu2} & U_{\mu3} \\
U_{\tau1} & U_{\tau2} & U_{\tau3}
\end{array}
\right)
\left(
\begin{array}{c}
\nu_1 \\
\nu_2 \\
\nu_3
\end{array}
\right) \nonumber \\
&=&
\left(
\begin{array}{ccc}
1 & 0 & 0 \\
0 & \cos\theta_{23} & \sin\theta_{23} \\
0 & -\sin\theta_{23} & \cos\theta_{23}
\end{array}
\right)
\left(
\begin{array}{ccc}
\cos\theta_{13} & 0 & \sin\theta_{13} e^{-i\delta} \\
0 & 1 & 0 \\
-\sin\theta_{13} e^{i\delta} & 0 & \cos\theta_{13}
\end{array}
\right)
\left(
\begin{array}{ccc}
\cos\theta_{12} & \sin\theta_{12} & 0 \\
-\sin\theta_{12} & \cos\theta_{12} & 0 \\
0 & 0 & 1
\end{array}
\right)
\left(
\begin{array}{c}
\nu_1 \\
\nu_2 \\
\nu_3
\end{array}
\right)~.
\end{eqnarray}
Using the above definition, neutrino flavor change can be described as a function of the mixing elements and neutrino masses in terms of the three-neutrino oscillation probability
\begin{eqnarray}
\label{3nu}
P(\nu_{\alpha}\to\nu_{\beta})=\delta_{\alpha\beta} &-& 4\sum_{i>j}\mathrm{Re}\{U^*_{\alpha i}U_{\alpha j}U_{\beta i}U^*_{\beta j}\}\sin^2\left(1.267\Delta m_{ij}^2L/E\right)  \nonumber \\
&+& \sum_{i>j}\mathrm{Im}\{U^*_{\alpha i}U_{\alpha j}U_{\beta i}U^*_{\beta j}\}\sin\left(2.534\Delta m_{ij}^2L/E\right)~,
\end{eqnarray}
where $\alpha,\beta=e,\mu,\tau$; $i,j=1,2,3$; $\Delta m^2_{ij}=m^2_i-m^2_j$ in eV$^2$; $E$ is the neutrino energy in MeV; and $L$ is the neutrino propagation distance (in the lab frame) from production to detection in meters. In a simpler, two-neutrino oscillation approximation, the oscillation probability is reduced to
\begin{eqnarray}
\label{2nu}
P(\stackrel{\tiny{(-)}}{\nu}_{\alpha}\to\stackrel{\tiny{(-)}}{\nu}_{\beta})=\delta_{\alpha\beta}-4|U^*_{\alpha 2}U_{\alpha 1}U_{\beta 2}U^*_{\beta 1}|\sin^2\left(1.267\Delta m^2_{21}L/E\right)~.
\end{eqnarray}
The above approximation holds when $\Delta m^2_{21}$ and $\Delta m^2_{31}$ differ significantly, and when only two of the three weak flavors ($\alpha$ and $\beta$) participate in the oscillation most-dominantly.

Neutrino oscillations described by the oscillation probabilities in Eqs.~\ref{3nu} and \ref{2nu} have now been established through multiple experiments \cite{pdg2011}. Those experiments study two-neutrino oscillations in the form of appearance of neutrinos of flavor $\beta$ in a neutrino beam of flavor $\alpha$, described by Eq.~\ref{2nu} when $\beta\ne\alpha$ ($\delta_{\alpha\beta}\equiv 0$), or in the form of disappearance of a neutrino beam of flavor $\alpha$, described by Eq.~\ref{2nu} when $\alpha\equiv\beta$ ($\delta_{\alpha\beta}\equiv1$).  

Almost all available experimental results are consistent with the following vacuum oscillation parameters \cite{pdg2011}:
\begin{eqnarray}
&\Delta m^2_{21}=m^2_2-m^2_1=7.65\times10^{-5}\mathrm{\ eV}^2~, \nonumber \\
&\sin^2\theta_{12}=0.304~, \nonumber \\
&\Delta m^2_{31}=m^2_3-m^2_1=2.40\times10^{-3}\mathrm{\ eV}^2~, \nonumber \\
&\sin^2\theta_{23}=0.5~, \nonumber \\
&0\le\sin^22\theta_{13}\le0.15~.
\end{eqnarray}
There are, however, existing results which cannot be accommodated within this picture, and suggest a possible need for extension beyond the three-neutrino framework. Those results come from both appearance and disappearance measurements performed at relatively short baselines. 

More specifically, two independent experiments, LSND \cite{Athanassopoulos:1996jb,Athanassopoulos:1997pv,Aguilar:2001ty} and MiniBooNE \cite{AguilarArevalo:2007it,AguilarArevalo:2008rc,AguilarArevalo:2009xn,AguilarArevalo:2010wv}, have observed three independent appearance-like excesses of electron neutrinos and/or antineutrinos in muon neutrino and/or antineutrino beams, at least two of which are consistent with oscillations at the level of 2.8-3.8 $\sigma$. Under a two-neutrino oscillation approximation, each of those two measurements reveal excesses which correspond to a large $\Delta m^2$, such that $\Delta m^2\gg\Delta m^2_{32}\gg\Delta m^2_{21}$, requiring at least one extra neutrino mass eigenstate to be added to the standard three-neutrino mass spectrum. This extra mass eigenstate is assumed to be mostly ``sterile'', i.e. non-weakly-interacting, by assumptions of unitarity and experimental constraints from $Z\rightarrow\nu\bar{\nu}$ decay measurements \cite{invZ}. Such models are referred to as 3 active + 1 sterile neutrino models, or, (3+1), and are usually explored in a two-neutrino oscillation approximation. In the case of MiniBooNE and/or LSND, this is done by setting $\Delta m^2_{32}\simeq\Delta m^2_{21}\simeq 0$ and $\Delta m^2_{41}\simeq\Delta m^2_{43}\simeq\Delta m^2_{42}$, so that the appearance oscillation probability can be obtained from Eq.~\ref{2nu}, for $\alpha=\mu$, $\beta=e$, and $\Delta m^2_{ij}=\Delta m^2_{41}$,
\begin{eqnarray}
\label{3p1app}
P(\stackrel{\tiny{(-)}}{\nu}_{\mu}\to\stackrel{\tiny{(-)}}{\nu}_e)=\sin^22\theta_{\mu e}\sin^2(1.267\Delta m^2_{41}L/E)~,
\end{eqnarray}
where we have used the orthogonality relation $U_{\mu1}U^*_{e1}=-U_{\mu4}U^*_{e4}$ and definition $\sin^22\theta_{\mu e}\equiv4|U_{e4}|^2|U_{\mu4}|^2$.

At the same time, a number of short-baseline reactor antineutrino experiments have looked for electron antineutrino disappearance at the same range of $\Delta m^2_{41}$, described by the $\bar{\nu}_e$ survival probability
\begin{eqnarray}
P(\stackrel{\tiny{(-)}}{\nu}_e\rightarrow\stackrel{\tiny{(-)}}{\nu}_e)&=&1-\sin^22\theta_{ee}\sin^2(1.27\Delta m^2_{41}L/E) \nonumber \\
&=&1-4|U_{e4}|^2(1-|U_{e4}|^2)\sin^2(1.27\Delta m^2_{41}L/E)~,
\end{eqnarray}
where we have replaced $|U_{e1}|^2$ with $(1-|U_{e4}|^2)$, by way of unitarity. While those same short-baseline reactor disappearance searches have provided strong limits in the past, a recent re-analysis of predicted reactor antineutrino fluxes revealed an underestimation of flux predictions previously assumed by those experiments \cite{Mueller}. As a result, the limits on $\sin^22\theta_{ee}$ have now been evaded, and an overall normalization reduction observed in their spectra shows consistency with $\sin^22\theta_{ee}\sim0.1$ and $\Delta m^2_{41}>$ 1 eV$^2$ \cite{reactor}.

It is tempting to attribute the MiniBooNE, LSND, and reactor short-baseline signals to the existence of a single sterile neutrino; however, attempted (3+1) fits have demonstrated that at least MiniBooNE neutrino results and LSND antineutrino results are incompatible under this scenario \cite{sterilewhite}, suggesting that any successful theoretical interpretation of those results must be more complex than simply (3+1). The source of this incompatibility is the fact that the MiniBooNE antineutrino mode and LSND antineutrino mode appearance searches prefer moderately small $\sin^22\theta_{\mu e}$ and $\Delta m^2\sim 1 $ eV$^2$, while the MiniBooNE neutrino mode appearance search disfavors such $\sin^22\theta_{\mu e}$ values for the same $\Delta m^2\sim1 $ eV$^2$.

A minimal extension to the (3+1) model, which would allow for CP-violation (i.e.~$P(\nu_{\mu}\rightarrow\nu_e)\ne P(\bar{\nu}_{\mu}\rightarrow\bar{\nu}_e)$), can successfully fit all of the above three signatures and seems plausible as an explanation. However, it still requires relatively large mixing amplitudes in order to reasonably accommodate them, and so it is consequently disfavored by $\nu_{\mu}$ disappearance experimental constraints \cite{Kopp:2011qd,Karagiorgi:2009nb,Giunti:2011gz}. 

In view of the shortcoming of the above, or ``vacuum,'' sterile neutrino oscillation models in reconciling MiniBooNE and LSND results, phenomenological efforts have now turned toward consideration of CPT-violating models \cite{Giunti:2010zs,Diaz:2011ia,AguilarArevalo:2011yi,Choudhury:2010vj,Barenboim:2009ts,everett}, or effectively CPT-violating models which involve non-standard matter effects \cite{Bramante:2011uu,Akhmedov:2010vy,Schwetz:2007cd,Nelson:2007yq}. Motivated by the latter class of models, in this paper, we consider a four-neutrino oscillation scenario, where the fourth neutrino flavor state, $\nu_s$, is subject to matter effects due to some interaction potential of the form
\begin{equation}
V_s=\pm A_s~,
\end{equation}
where $A_s$ is a constant and $+/-$ corresponds to neutrino/antineutrino propagation through matter. This generalization allows us to follow an agnostic approach as to the underlying source of this effect. We assume that the excesses observed by MiniBooNE and LSND are manifestations of $\nu_{\mu}\rightarrow\nu_e$ oscillation via the fourth mass eigenstate, which is assumed to be on the order of 0.01-100~eV$^2$, and a mix of $\nu_e$ and $\nu_{\mu}$ eigenstates at the $\le5$\% level each ($|U_{e4}|^2,|U_{\mu4}|^2\le0.05$), and $\nu_s$ at the $\ge90$\% level ($|U_{s4}|^2\ge0.90$).

We discuss the oscillation framework in detail in the following section. In 
Sec.~\ref{sec:four} we describe the analysis and fit machinery used to apply this framework to the MiniBooNE and LSND data sets. In Sec.~\ref{sec:five}, we present quantitative results, first in a (3+1) scenario without this matter effect ($A_s=0$), as a reference, and then with the matter effect turned on. In Sec.~\ref{sec:five}, we provide a qualitative description of the results, and, in Sec.~\ref{sec:six}, we discuss implications for atmospheric, solar, and reactor experiments. Conclusions are presented in Sec.~\ref{sec:seven}.

\section{\label{sec:two}OSCILLATION FRAMEWORK}

In this section, we derive the $\nu_{\mu}\rightarrow\nu_e$ and $\bar{\nu}_{\mu}\rightarrow\bar{\nu}_e$ appearance oscillation probabilities to which we attribute the observed MiniBooNE and LSND excesses. For simplicity, we assume that the matter potential $V_s$ experienced by $\nu_s$ is much larger in amplitude than the standard model matter effect potentials, $V_{CC}$ and $V_{NC}$, so that the effective matter potential in neutrino flavor space can be approximated as
\begin{eqnarray}
V=\left(\begin{array}{cccc}
V_{CC}+V_{NC} & 0 & 0 & 0 \\
0 & V_{NC} & 0 & 0 \\
0 & 0 & V_{NC} & 0 \\
0 & 0 & 0 & V_s
\end{array}
\right)
\simeq
\left(
\begin{array}{cccc}
0 & 0 & 0 & 0 \\
0 & 0 & 0 & 0 \\
0 & 0 & 0 & 0 \\
0 & 0 & 0 & V_s \\
\end{array}\right)~.
\end{eqnarray}

The effective Hamiltonian for neutrino propagation in matter, expressed in flavor space, is given by
\begin{eqnarray}
\label{hflavor}
H_m=\frac{1}{2E}U\left(\begin{array}{cccc}
0 & 0 & 0 & 0 \\
0 & 0 & 0 & 0 \\
0 & 0 & 0 & 0 \\
0 & 0 & 0 & m_4^2
\end{array}
\right)U^\dagger + V~.
\end{eqnarray}
For simplicity, we have assumed that $m^2_1$, $m^2_2$ and $m^2_3$ in vacuum are degenerate and negligible relative to $m^2_4$, with $\Delta m^2_{41}\equiv\Delta m^2$ being the only mass squared difference in vacuum. We also assume that $m^2_{2}/2E$ and $m^2_{3}/2E$ are negligible relative to $V_s$. Then, using the standard form of the $U$ mixing matrix in vacuum,
\begin{eqnarray}
U=\left(\begin{array}{cccc}
U_{e1} & U_{e2} & U_{e3} & U_{e4} \\
U_{\mu1} & U_{\mu2} & U_{\mu3} & U_{\mu4} \\
U_{\tau1} & U_{\tau2} & U_{\tau3} & U_{\tau4} \\
U_{s1} & U_{s2} & U_{s3} & U_{s4} \\
\end{array}
\right)~,
\end{eqnarray}
the effective Hamiltonian in matter becomes
\begin{eqnarray}
\label{hflavor2}
H_m=\frac{\Delta m^2}{2E}
\left(\begin{array}{cccc}
U_{e4}U^*_{e4} &  U_{e4} U^*_{\mu4} &  U_{e4}U_{\tau4}^* & U_{e4}U_{s4}^* \\
U_{\mu4}U_{e4}^* &  U_{\mu4}U^*_{\mu4} &  U_{\mu4}U_{\tau4}^* & U_{\mu4}U_{s4}^* \\
U_{\tau4}U_{e4}^* & U_{\tau4}U_{\mu4}^* & U_{\tau4}U^*_{\tau4} & U_{\tau4}U_{s4}^* \\
U_{s4}U_{e4}^* & U_{s4}U_{\mu4}^* & U_{s4}U_{\tau4}^* & U_{s4}U^*_{s4}+2EV_s/\Delta m^2
\end{array}
\right)~.
\end{eqnarray}
Diagonalizing the effective Hamiltonian in Eq.~\ref{hflavor2} gives the following eigenvalues:
\begin{eqnarray}
&\lambda_1 = 0~, \nonumber \\
&\lambda_2 = 0~, \nonumber \\
&\lambda_3 = \frac{1}{4E}\left(2EV_s+\Delta m^2\sum_{\alpha}|U_{\alpha4}|^2 -
\sqrt{(-2EV_s-\Delta m^2\sum_{\alpha}|U_{\alpha4}|^2)^2-8EV_s\Delta m^2(\sum_{\alpha}|U_{\alpha4}|^2-|U_{s4}|^2)}\right)~, \nonumber \\ 
&\lambda_4 = \frac{1}{4E}\left(2EV_s+\Delta m^2\sum_{\alpha}|U_{\alpha4}|^2 +
\sqrt{(-2EV_s-\Delta m^2\sum_{\alpha}|U_{\alpha4}|^2)^2-8EV_s\Delta m^2(\sum_{\alpha}|U_{\alpha4}|^2-|U_{s4}|^2)}\right)~,
\end{eqnarray}
which, by unitarity, are reduced to
\begin{eqnarray}
&\lambda_1=0~,\nonumber \\
&\lambda_2=0~,\nonumber \\
&\lambda_3 = \frac{1}{4E}\left(2EV_s+\Delta m^2 -
\sqrt{(2EV_s+\Delta m^2)^2-8EV_s\Delta m^2(1-|U_{s4}|^2)}\right)~,\nonumber \\ 
&\lambda_4 = \frac{1}{4E}\left(2EV_s+\Delta m^2 +
\sqrt{(2EV_s+\Delta m^2)^2-8EV_s\Delta m^2(1-|U_{s4}|^2)}\right)~.
\end{eqnarray}
The differences $\lambda_4-\lambda_{1,2}$, $\lambda_4-\lambda_3$, and $\lambda_3-\lambda_{1,2}$ suggest three distinct effective $\Delta m^2_M$ values in matter. Note, however, that, if the active flavor content of the fourth mass eigenstate is small, then we can approximate 
\begin{equation}
\label{approximation}
1-|U_{s4}|^2\simeq0~, 
\end{equation}
in which case all eigenvalues except $\lambda_{4}$ become zero. The non-zero $\lambda_4$,
\begin{equation}
\lambda_4=\frac{1}{2E}(2EV_s+\Delta m^2)~,
\end{equation} 
implies one effective $\Delta m^2_M$, which corresponds to 
\begin{equation}
\label{effdm2}
\Delta m^2_M=\Delta m^2 + 2EV_s~.
\end{equation}
We will be using the above $\Delta m^2_M$ approximation in our fits, which is a justified assumption according to the level of unitarity in the three-neutrino mixing matrix, which is experimentally constrained \cite{gkthesis,Antusch:2007zza,Antusch:2006vwa}. Note that, when $V_s\rightarrow0$, $\Delta m^2_M$ reduces to the vacuum $\Delta m^2$ value, as expected.

The matrix consisting of the unit-normalized eigenvectors (columns) of the effective Hamiltonian in Eq.~\ref{hflavor2}, defines the new mixing matrix in matter, and respective mixing elements,
\begin{equation}
U^M=P~.
\end{equation}

For the purposes of this paper, we are interested in the general expression from which the $\nu_{\mu}\rightarrow\nu_e$ and $\bar{\nu}_{\mu}\rightarrow\bar{\nu}_e$ appearance oscillation probabilities are derived,
\begin{eqnarray}
\label{stdoscprob}
P(\nu_{\mu}\rightarrow\nu_e)= |\sum_iU^*_{ei}e^{-im^2_iL/2E}U_{\mu i}|^2~.
\end{eqnarray}
According to Eq.~\ref{stdoscprob}, the elements of interest to this appearance channel are $U^M_{ei}$ and $U^M_{\mu i}$. Since only one $\Delta m^2_M$ dominates, that associated with $\lambda_4$, the parameters of interest are just $(U^M_{e4})^*$ and $U^M_{\mu4}$. From the eigenvectors of $H_m$, one identifies
\begin{eqnarray}
\label{effu4}
U^M_{e4}=&\Delta m^2 U_{e4}U^*_{s4}\sqrt{\frac{1+4\Delta m^2|U_{s4}|^2(1-|U_{s4}|^2)/\left(2EV_s-\Delta m^2(1-2|U_{s4}|^2)+\sqrt{(\Delta m^2-2EV_s)^2+8EV_s\Delta m^2|U_{s4}|^2}\right)^2}{(\Delta m^2-2EV_s)^2+8EV_s\Delta m^2|U_{s4}|^2}}~, \nonumber \\
&
\end{eqnarray}
and
\begin{eqnarray}
\label{effum4}
U^M_{\mu4}=&\Delta m^2 U_{\mu4}U^*_{s4}\sqrt{\frac{1+4\Delta m^2|U_{s4}|^2(1-|U_{s4}|^2)/\left(2EV_s-\Delta m^2(1-2|U_{s4}|^2)+\sqrt{(\Delta m^2-2EV_s)^2+8EV_s\Delta m^2|U_{s4}|^2}\right)^2}{(\Delta m^2-2EV_s)^2+8EV_s\Delta m^2|U_{s4}|^2}}~. \nonumber \\
\end{eqnarray}
The oscillation probability is then derived just as in the standard (3+1) neutrino oscillation scenario (from Eq.~\ref{stdoscprob}),
\begin{equation}
\label{mfxprob}
P(\nu_{\mu}\rightarrow\nu_e)=4|U^M_{e4}|^2|U^M_{\mu4}|^2\sin^2(1.27\Delta m^2_{M}L/E)~,
\end{equation}
where one has replaced $\Delta m^2$ with $\Delta m^2_M$ from Eq.~\ref{effdm2} and $U_{\alpha 4}$ with $U^M_{\alpha 4}$ from Eqs.~\ref{effu4} and~\ref{effum4}. The effective mixing amplitude in matter, $\sin^22\theta^M_{\mu e}=4|U^M_{e4}|^2|U^M_{\mu4}|^2$, can be expressed in terms of $|U_{e4}|$ and $|U_{\mu4}|$ as
\begin{eqnarray}
\label{full}
\sin^22\theta^M_{\mu e}=&\frac{16(\Delta m^2)^4 |U_{e4}|^2|U_{\mu4}|^2|U_{s4}|^4}{\left((\Delta m^2-2EV_s)^2+8EV_s\Delta m^2|U_{s4}|^2\right)\left(2EV_s-\Delta m^2(1-2|U_{s4}|^2)+\sqrt{(\Delta m^2-2EV_s)^2+8EV_s\Delta m^2|U_{s4}|^2}\right)^2}~. \nonumber \\
\end{eqnarray}
Note that $\sin^22\theta^M_{\mu e}$ reduces to $\sin^22\theta_{\mu e}=4|U_{e4}|^2|U_{\mu4}|^2$ when $V_s\to0$, as expected.

As in the case of standard matter effects, resonances occur when the denominator of Eq.~\ref{full} is minimized \footnote{Note that the approximation in Eq.~\ref{approximation} is not applied in the $\sin^22\theta_{\mu e}^M$ expression.}. It is clear that, because of the change in sign in $V_s$ for neutrinos versus antineutrinos, we generally expect resonances to appear at different energies for neutrinos versus antineutrinos, even when the measurements are performed at the same $L$ and $E$, while underlying vacuum oscillation parameters, $\Delta m^2$, $|U_{e4}|$, $|U_{\mu4}|$ and $|U_{s4}|$, are assumed to be the same for neutrinos and antineutrinos, as expected by CPT conservation. Upon inspection of the denominator in Eq.~\ref{full}, given the $V_s=\pm A_s$ definition for neutrinos/antineutrinos, one expects resonances to occur in the case of antineutrino but not neutrino oscillations.

\begin{figure}[t]
\begin{center}
\includegraphics[width=4in, angle=-90, trim=50 100 50 0]{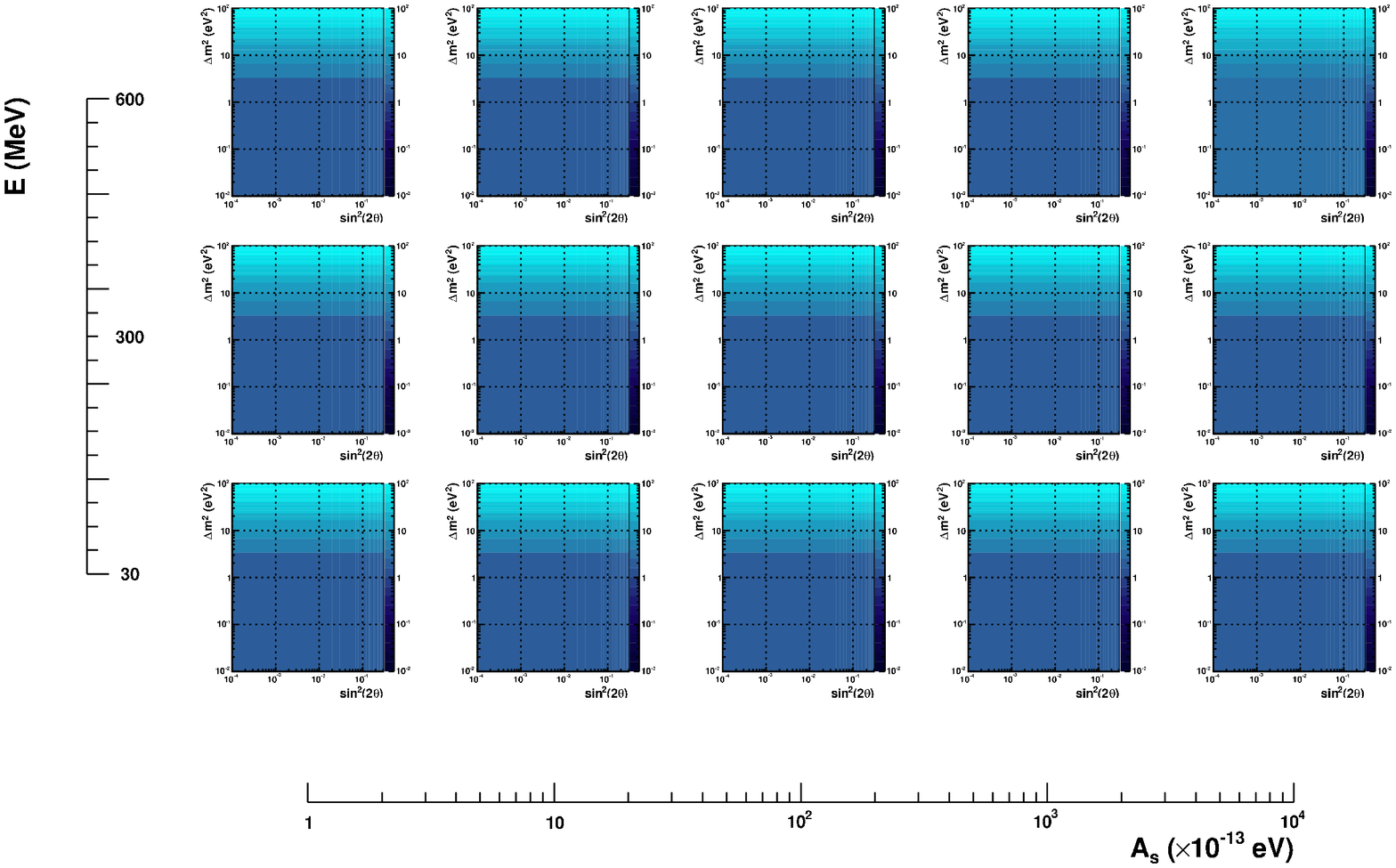}
\includegraphics[width=4in, angle=-90, trim=50 100 50 0]{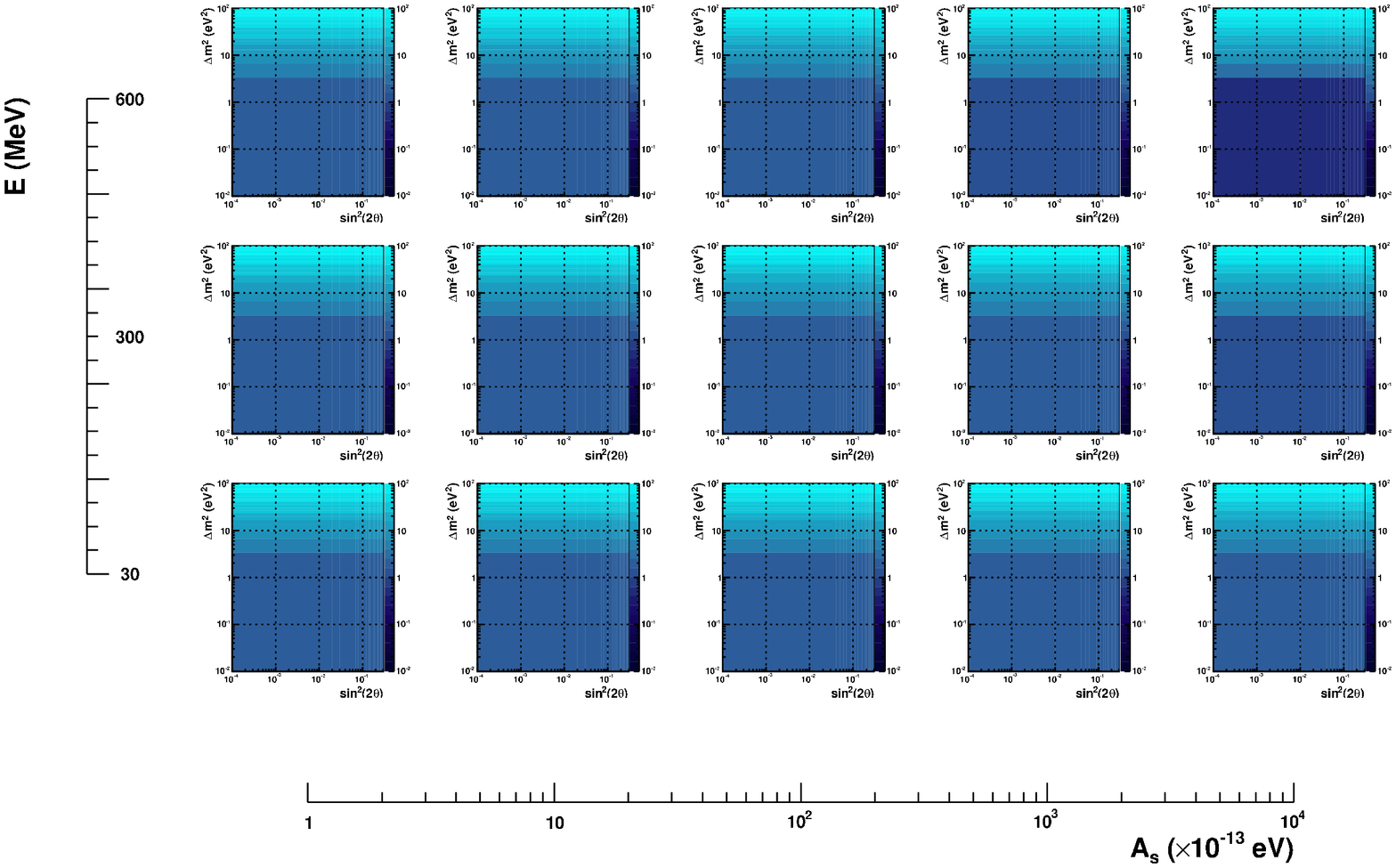}
\end{center}
\caption{\label{giant1}Effective $\Delta m^2_M$ (colored contours, with color scale on the right) as a function of the vacuum oscillation parameters $\Delta m^2$ and $\sin^22\theta_{\mu e}$ for different sets of neutrino energy values $E$ and matter effect potential $A_s$. The top set of plots corresponds to neutrinos. The bottom set corresponds to antineutrinos.}
\end{figure}

\begin{figure}[t]
\begin{center}
\includegraphics[width=4in, angle=-90, trim=50 100 50 0]{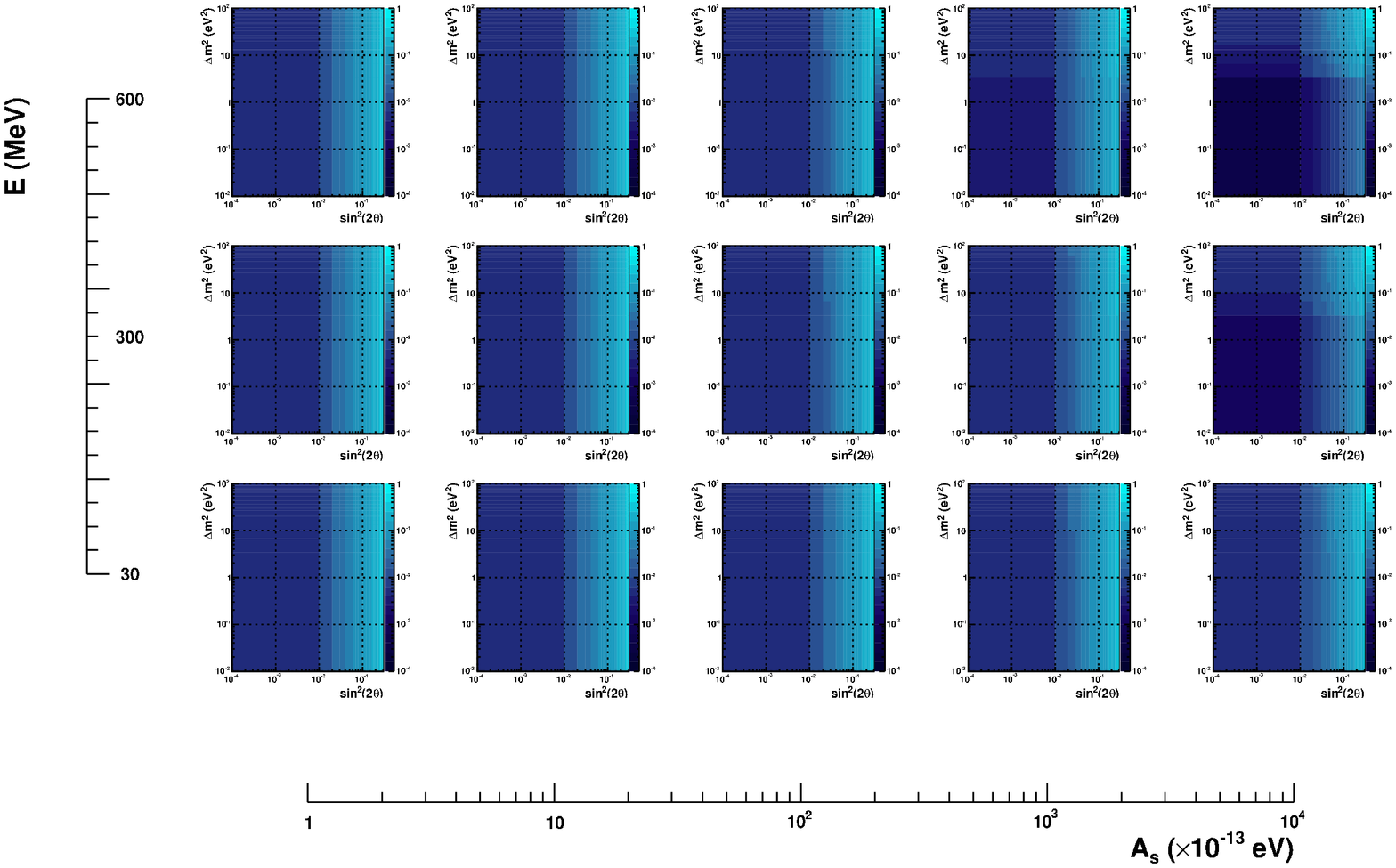}
\includegraphics[width=4in, angle=-90, trim=50 100 50 0]{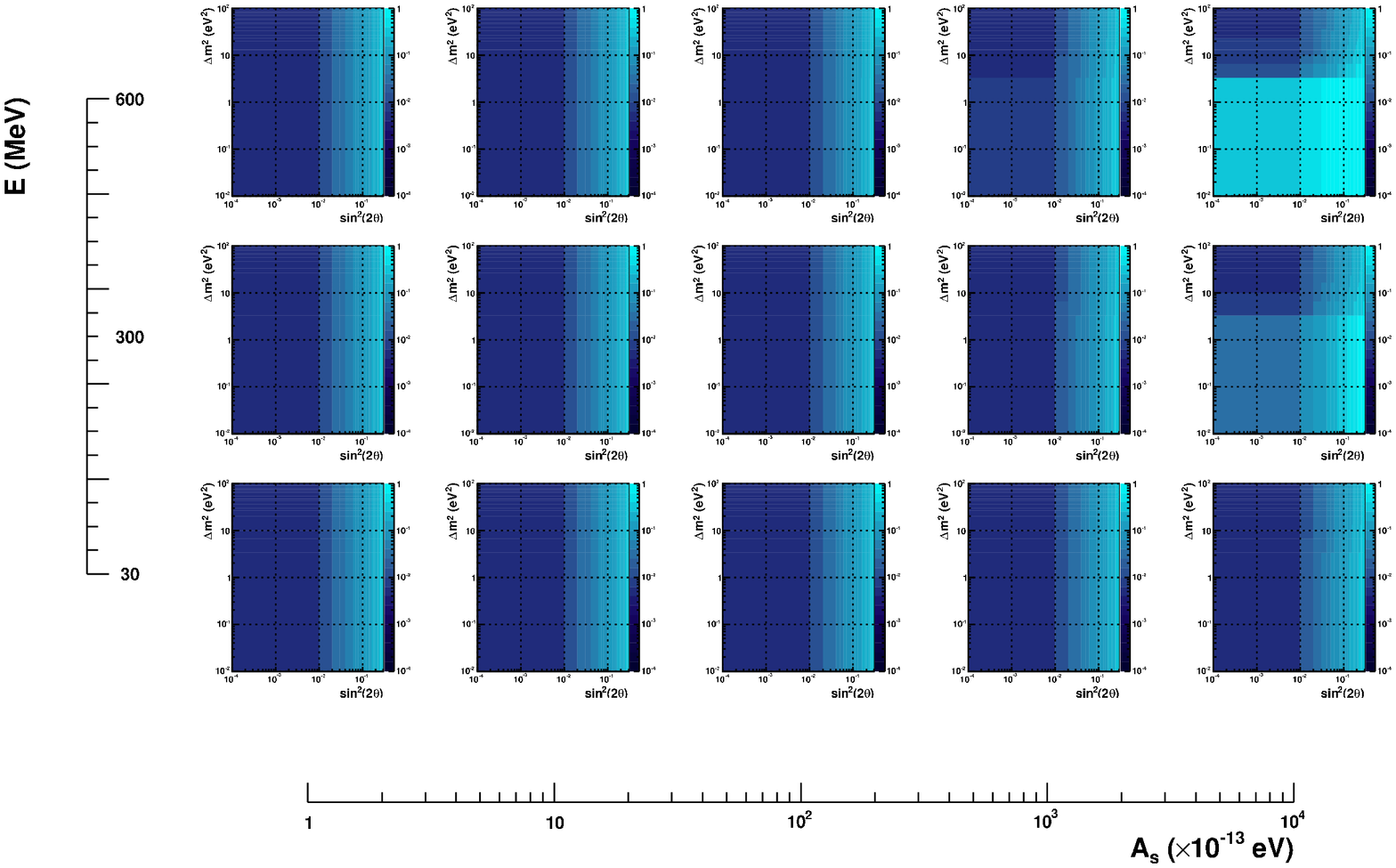}
\end{center}
\caption{\label{giant2}Effective $\sin^22\theta_{\mu e}^M$ (colored contours, with color scale on the right) as a function of the vacuum oscillation parameters $\Delta m^2$ and $\sin^22\theta_{\mu e}$ for different sets of neutrino energy values $E$ and matter effect potential $A_s$. The top set of plots corresponds to neutrinos. The bottom set corresponds to antineutrinos.}
\end{figure}

Because $\sin^22\theta^M_{\mu e}$ and $\Delta m^2_M$ are the parameters measured by experiments by fitting to the usual two-neutrino appearance probability formula, given in Eq.~\ref{stdoscprob}, it is instructive to examine $\sin^22\theta^M_{\mu e}$ and $\Delta m^2_M$ as a function of $E$, $A_s$, $\sin^22\theta_{\mu e}$, and $\Delta m^2$. Figures~\ref{giant1} and \ref{giant2} illustrate the dependence of $\Delta m^2_M$ (Fig.~\ref{giant1}) and $\sin^22\theta^M_{\mu e}$ (Fig.~\ref{giant2}) on the underlying vacuum oscillation parameters, $\sin^22\theta_{\mu e}$ and $\Delta m^2$, for specific values of $E$ and $A_s$. We have purposefully picked neutrino energies $E$ close to the LSND, MiniBooNE low energy and MiniBooNE high energy excess mean energies. The $A_s$ values have been chosen to span the orders of magnitude considered in our fit. The top set of plots in each figure illustrates this dependence for the case of neutrinos, while the bottom set illustrates the antineutrino case. From the figures, one can see that as $A_s\rightarrow0$, the effective $\sin^22\theta^M_{\mu e}$ and $\Delta m^2_M$ approach the vacuum $\sin^22\theta_{\mu e}$ and $\Delta m^2$ parameters. However, as $A_s$ turns on, they deviate from the vacuum values, and increasingly so with $E$. Resonances can be identified in lighter regions, where $\sin^22\theta_{\mu e}^M\rightarrow1$.

%
%

\section{\label{sec:four}ANALYSIS METHOD}

We assume that the LSND and MiniBooNE excesses can be described by the oscillation probability in Eq.~\ref{mfxprob}, and we fit for the vacuum oscillation parameters $|U_{e4}|$, $|U_{\mu 4}|$ and $\Delta m^2$, and $A_s$, on which $\Delta m^2_M$ and $\sin^22\theta^M_{\mu e}$ depend. We set $|U_{s4}|^2=1-|U_{e4}|^2-|U_{\mu4}|^2$. The vacuum oscillation parameters are allowed to vary freely within $10^{-3}<\Delta m^2<100$ eV$^2$ and $\sin^22\theta_{\mu e}<0.01$. The $\sin^22\theta_{\mu e}$ upper bound is limited by requiring $|U_{e4}|^2<0.05$ and $|U_{\mu4}|^2<0.05$, assuming unitarity of the 3$\times$3 matrix at the 5\% level. The matter potential is allowed to vary freely within $10^{-13}<A_s<10^{-9}$ eV. The range of $A_s$ chosen in our fits has been mainly motivated by the assumption that this new potential should lead to observable effects at $L/E\sim$~1~eV$^2$, which is supported by the findings in \cite{Kamo:2002sj}. For comparison, the standard matter effect pottential considered in long-baseline neutrino oscillation experiments is $\sqrt{2}G_Fn_e\sim10^{-13}$ eV. 

During the fit, the model parameters $|U_{\mu4}|$, $|U_{e4}|$, $\Delta m^2$, and $A_s$ are generated and varied according to a Markov Chain \cite{braemaud,Metropolis:1953am} $\chi^2$ minimization routine. For each variation, the signal predictions for MiniBooNE neutrino mode, MiniBooNE antineutrino mode and LSND are calculated using the oscillation probability in Eq.~\ref{mfxprob} and then compared to the observed excesses in the form of a $\chi^2$. 

The LSND and MiniBooNE data sets used in fits presented in this paper are identical to those in \cite{Karagiorgi:2009nb}, with the exception of the MiniBooNE antineutrino data set, where we use the higher-statistics, updated results from the MiniBooNE $\bar{\nu}_{\mu}\rightarrow\bar{\nu}_e$ search, corresponding to 5.66$\times$10$^{20}$ protons on target (POT) \cite{AguilarArevalo:2010wv}. Note that MiniBooNE antineutrino data taking is still in progress. The experiment aims to complete its total antineutrino running at the end of this spring, with an estimated final antineutrino sample corresponding to $\sim$10$\times$10$^{20}$ POT. The fits presented in this paper should therefore be updated once the higher statistics antineutrino data set from MiniBooNE becomes available.

\indent The MiniBooNE neutrino data set (MB$\nu$) is included in the fits in the form of two side-by-side distributions of $\nu_e$ and $\nu_{\mu}$ charged-current quasi-elastic (CCQE) candidate events, each as a function of neutrino energy which has been reconstructed assuming CCQE scattering. Using both the $\nu_{\mu}$ and $\nu_e$ CCQE reconstructed events in the fit along with their correlations significantly reduces some of the systematic errors related to neutrino flux and cross section. The full 200-3000 MeV range of $\nu_e$ CCQE data available is used in the fit. The observed event distributions are compared to the corresponding Monte Carlo predicted distributions, and a $\chi^2$ is calculated using a covariance matrix which includes systematic and statistical uncertainties as well as systematic correlations between the predicted $\nu_e$ and $\nu_{\mu}$ distributions. During the fit, we vary the $\nu_e$ distribution according to the sterile neutrino vacuum oscillation parameters and $A_s$, but we keep the $\nu_{\mu}$ distribution unchanged, despite the possibility of $\nu_{\mu}$ disappearance in the MiniBooNE data. We verify that the best-fit model implies $\nu_{\mu}$ disappearance well below the level constrained by the MiniBooNE, SciBooNE, and MINOS $\nu_{\mu}$ disappearance analyses \cite{AguilarArevalo:2009yj,Mahn:2011ea,Adamson:2011ku}, and assume that the effect of any $\nu_{\mu}$ disappearance on the ability of the $\nu_{\mu}$ CCQE sample to constrain the $\nu_e$ CCQE sample is small. The fit method follows the details described in \cite{Karagiorgi:2009nb}.

\indent The MiniBooNE antineutrino data set (MB$\bar{\nu}$) is included in the fits in the same way as the MB$\nu$ data set, in the form of two side-by-side distributions of $\bar{\nu}_e$ and $\bar{\nu}_{\mu}$ CCQE candidate events. The full 200-3000 MeV range of $\bar{\nu}_e$ CCQE data is used in the fit. The MB$\bar{\nu}$ data fit method also follows the details described in \cite{Karagiorgi:2009nb}.

\indent The LSND data set is included in the fits in the form of a $\bar{\nu}_e$ event distribution from $\bar{\nu}_ep\rightarrow ne^+$ interactions, as a function of five positron energy bins between 20 and 60~MeV. We neglect the higher-energy pion-decay-in-flight sample and fit only the decay-at-rest sample which makes this a pure $\bar{\nu}_{\mu}\rightarrow\bar{\nu}_e$ search. Because this is a low-statistics sample, the LSND $\chi^2$ function is constructed as a log-likelihood function of the observed data, expected signal, and expected backgrounds. The fit method follows the details of \cite{sorel,karagiorgi1,Karagiorgi:2009nb}.

The $\chi^2$ returned by each data set is used to construct the total $\chi^2$, 
\begin{equation}
\chi^2_{total} = \chi^2_{MB\nu}+\chi^2_{MB\bar{\nu}}+\chi^2_{LSND}~,
\end{equation}
which is minimized in the Markov Chain, and then used to extract the best fit and 90\% and 99\% confidence level (CL) allowed oscillation parameter regions. The CL intervals shown in the figures of this paper correspond to the standard $\Delta\chi^2$ cuts for two (2) degrees of freedom.

In order to quantify the statistical compatibility between various data sets under a particular hypothesis, we use the Parameter Goodness-of-fit (PG) test introduced in \cite{Maltoni:2003cu}. This test quantifies the level of agreement between various data sets by comparing the minimum $\chi^2$ obtained by a simultaneous fit to all data sets, $\chi^2_{min,all}$, to the sum of the individual minimum $\chi^2$'s obtained by a separate fit to each of the data sets, i.e.,
\begin{equation}
\chi^2_{PG}=\chi^2_{min,all} - \sum_{i}\chi^2_{min,i}~,
\end{equation}  
where $i$ runs over the data sets considered in the fit yielding $\chi^2_{min,all}$. The PG is obtained from $\chi^2_{PG}$ based on the number of common underlying fit parameters, $ndf_{PG}$, using the standard probability distribution function.


\section{\label{sec:five}RESULTS}

\subsection{\label{sec:fivea}Fit results with no matter effects: $A_s=0$}

\indent In this section, the MiniBooNE and LSND results are examined under a (3+1) oscillation hypothesis with no matter effects ($A_s=0$). The results are to be used as a reference in Secs.~\ref{sec:fiveb} and \ref{sec:fivec}. Note that results presented in this section differ from those reported in \cite{Karagiorgi:2009nb} because of (a) the updated MB$\bar{\nu}$ data set being used here and (b) the $|U_{e4}|^2\le0.05$ and $|U_{\mu4}|^2\le0.05$ fit constraints imposed in this analysis.

A (3+1) fit to oscillations is unable to reconcile the three signatures. The $\chi^2$-probability for the best-fit parameters obtained from a simultaneous fit to all three data sets corresponds to 6.8\% ($\chi^2/ndf=52.89/39$). The compatibility between all three data sets is found to be 2.3\%, using the PG criterion 
\begin{equation}
PG=prob(\chi^2_{PG},ndf_{PG})=(11.4,4)=2.3\mathrm{\%}~.
\end{equation}

\begin{figure}[h]
\begin{center}
\includegraphics[width=3in, trim=0 0 0 0]{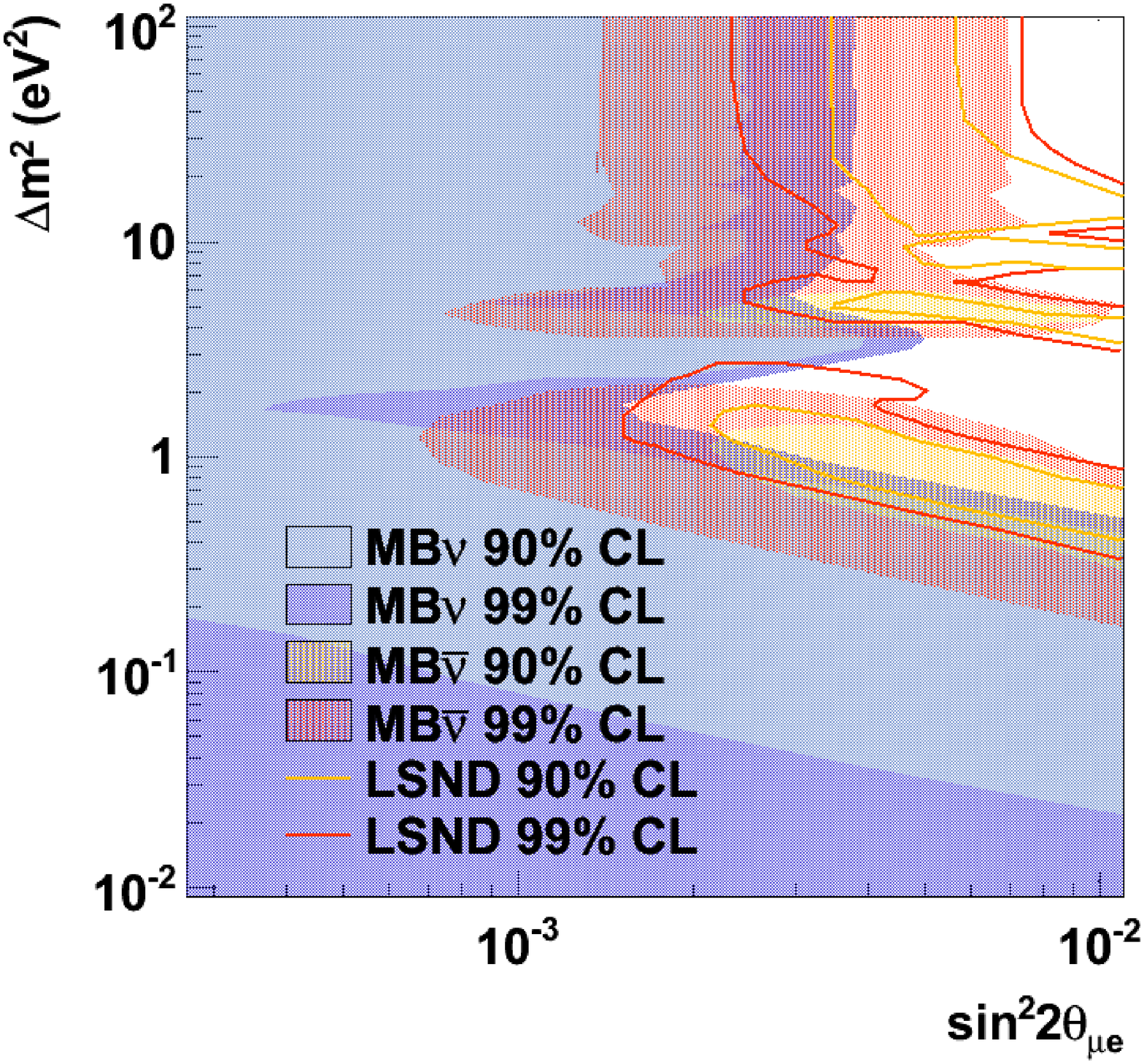}
\end{center}
\caption{\label{3p1_overlay}{90\% and 99\% confidence level (CL) allowed regions from fitting each of the data sets (MB$\nu$, MB$\bar{\nu}$ and LSND) seperately to a (3+1) model. While MB$\bar{\nu}$ and LSND allowed regions overlap in regions of high confidence level, the MB$\nu$ allowed regions are significantly shifted to lower $\sin^22\theta_{\mu e}$ values.}}
\end{figure}

\begin{figure}[h]
\begin{center}
\includegraphics[width=4in, trim=0 0 0 0]{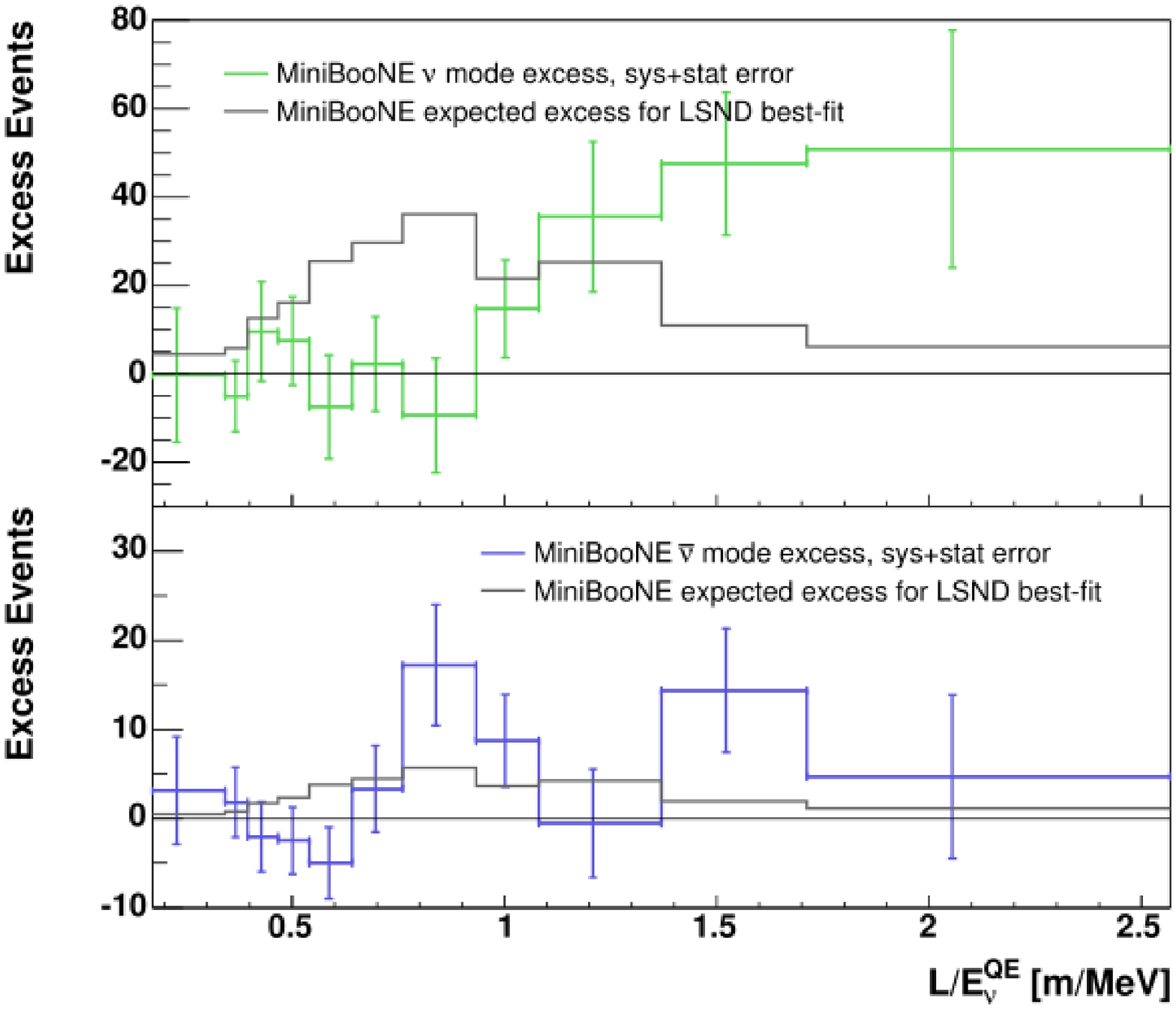}
\end{center}
\caption{\label{LE}The MiniBooNE neutrino (top) and antineutrino (bottom) observed excess distributions each as a function of $L/E$, $E$ being the reconstructed neutrino energy, $E^{QE}_{\nu}$. The error bars include systematic and statistical uncertainties. Overlaid are signal predictions corresponding to the LSND (3+1) best-fit parameters. The figure is taken from \cite{gkthesis}.}
\end{figure}

The source of incompatibility is demonstrated in Fig.~\ref{3p1_overlay}. While MiniBooNE antineutrino and LSND (antineutrino) results yield contours which overlap in regions of high confidence level, the MiniBooNE neutrino results highly exclude those regions, and are preferentially shifted to lower ($\sin^22\theta_{\mu e}$,$\Delta m^2$) values. The incompatibility is also illustrated in Fig.~\ref{LE}, where the phase of the MiniBooNE neutrino observed $L/E$ distribution is significantly shifted relative to that of the MiniBooNE antineutrino data set or that expected from the LSND best-fit prediction.

\subsection{\label{sec:fiveb}Fit results with matter effects: $A_s\ne0$}

\indent In this section, the MiniBooNE and LSND results are examined under a (3+1) oscillation hypothesis with matter effects ($A_s\ne0$). The $\chi^2$-probability for the best-fit parameters obtained from a simultaneous fit to all three data sets corresponds to 21.6\% ($\chi^2/ndf=44.54/38$). Relative to the (3+1) fit in Sec.~\ref{sec:fivea}, the best fit $\chi^2$ is reduced by 8.35 units for one extra fit parameter. The compatibility between all three data sets is found to be 17.4\%, using the PG criterion
\begin{equation}
PG=prob(\chi^2_{PG},ndf_{PG})=(9.0,6)=17.4\mathrm{\%}~.
\end{equation}
The $\chi^2$ probability for each experiment, at the best-fit parameters found by the joint fit, corresponds to 25.0\% (MB$\nu$), 15.3\% (MB$\bar{\nu}$) and 12.3\% (LSND), supporting the high compatibility reported above.

\begin{figure}[t]
\begin{center}
\includegraphics[width=3.5in, trim=0 0 0 0]{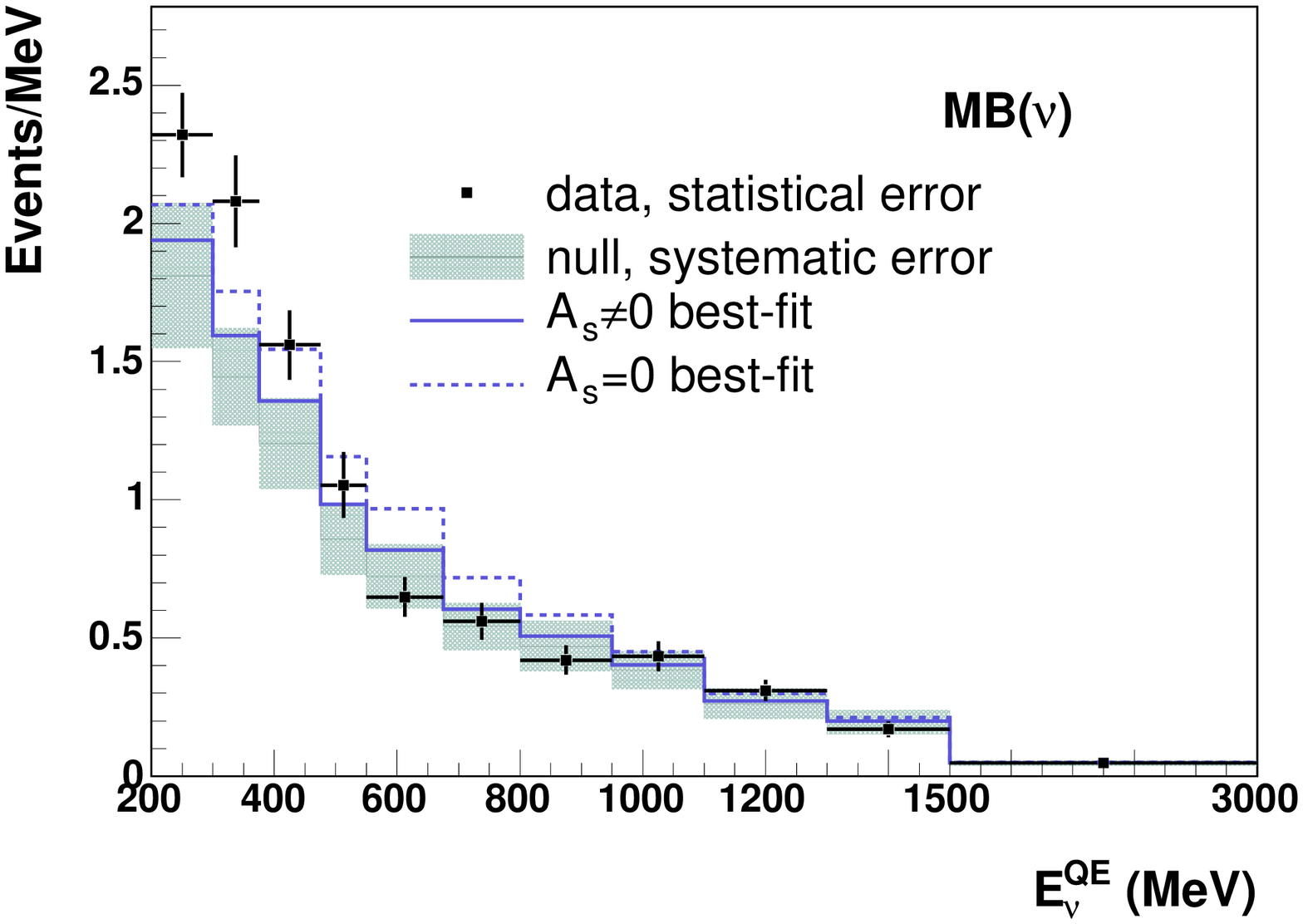}\\
\includegraphics[width=3.5in, trim=0 0 0 0]{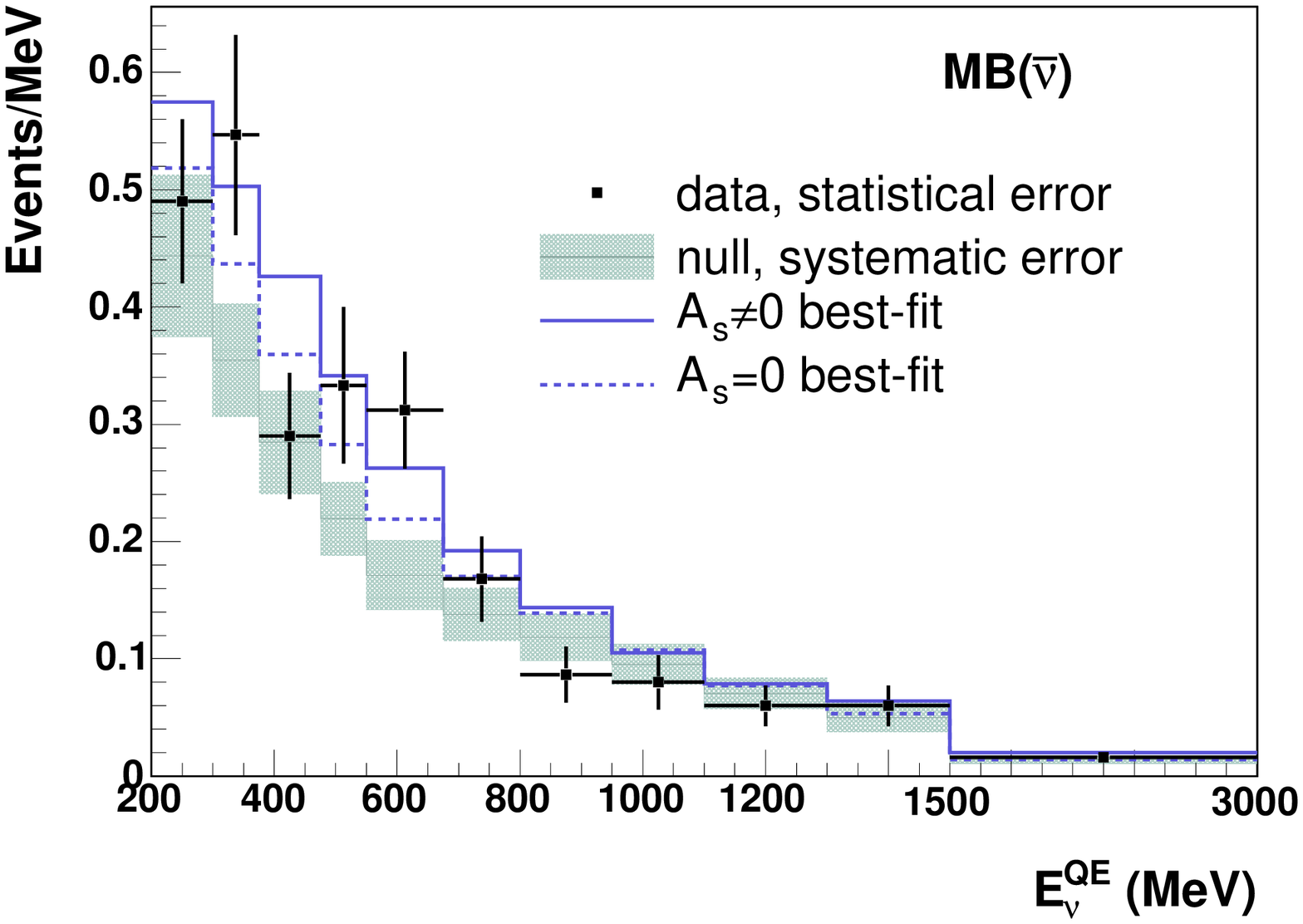}\\
\includegraphics[width=3.5in, trim=0 20 0 -20]{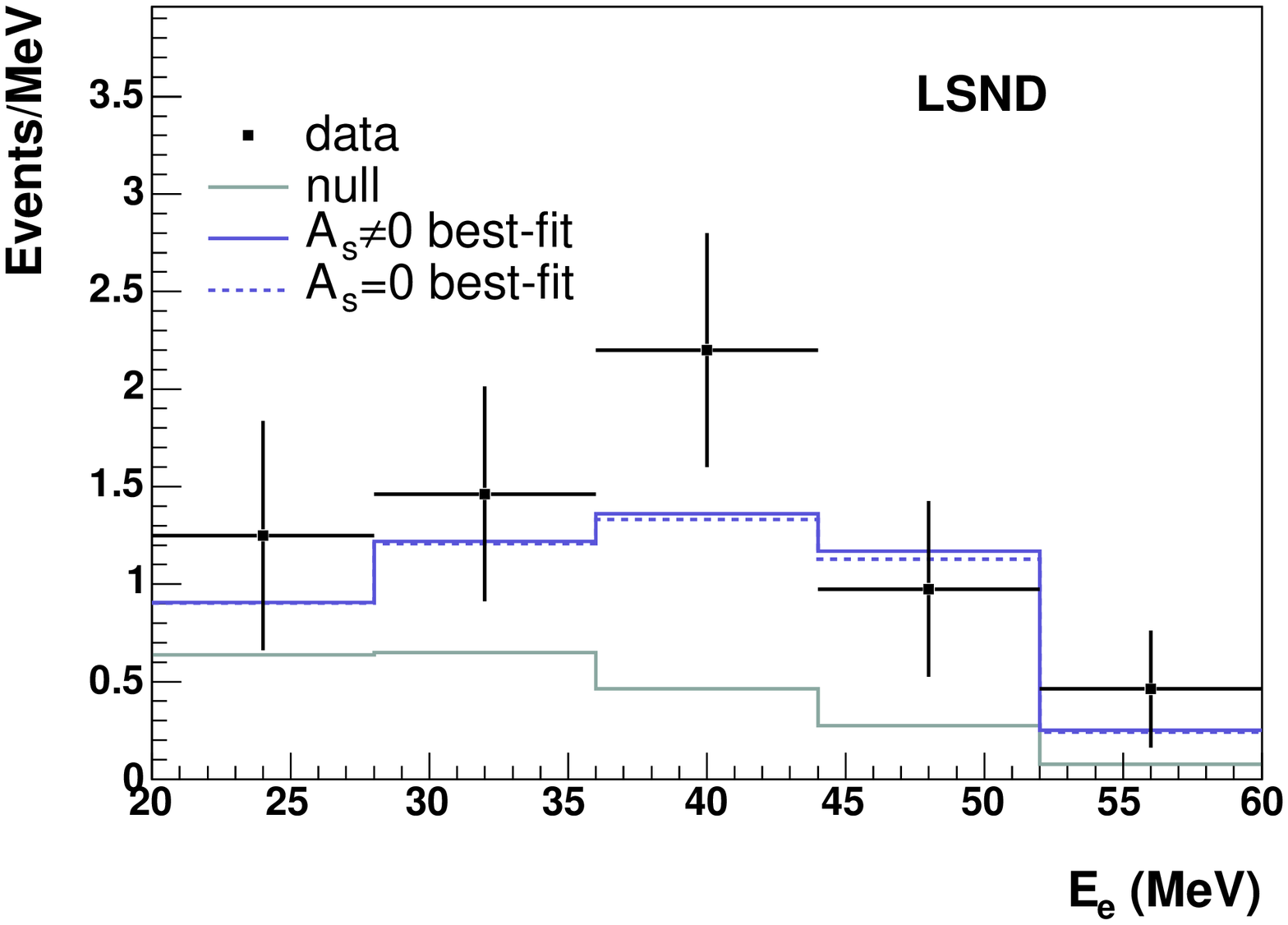}
\end{center}
\caption{\label{distributions}Best-fit distributions for the matter effects fit ($A_s\ne0$), in solid blue, and the (3+1) fit ($A_s=0$), in dashed blue, overlaid on the MiniBooNE neutrino (top), antineutrino (middle), and LSND (bottom) observed spectra. The no oscillations prediction is also shown, in gray. In the case of the MiniBooNE neutrino distributions, the matter effect fit predicts a lower excess at low energy, but also no excess at higher energy, which reduces the tension otherwise present in the (3+1) fit. In the case of the MiniBooNE antineutrino distributions, the matter effect fit predicts both low and high energy excess. In the case of the LSND distributions, the two fits are essentially indistinguishable.}
\end{figure}

The best-fit signal plus background predictions as a function of energy are shown in Fig.~\ref{distributions}, for each of the three data sets. For comparison, the (3+1) best-fit predictions are also overlaid. The (3+1) fit with matter effects accommodates a smaller fraction of the low energy excess observed in MiniBooNE neutrino mode than the (3+1) fit, but describes the high energy data significantly better. In antineutrino mode the fit predicts a considerably larger excess than the (3+1) fit, across all energies. In the case of LSND, the matter effect fit performs only marginally better than the (3+1) fit.

\begin{figure}[t]
\begin{center}
\includegraphics[width=3.5in, angle=-90, trim=0 0 0 0]{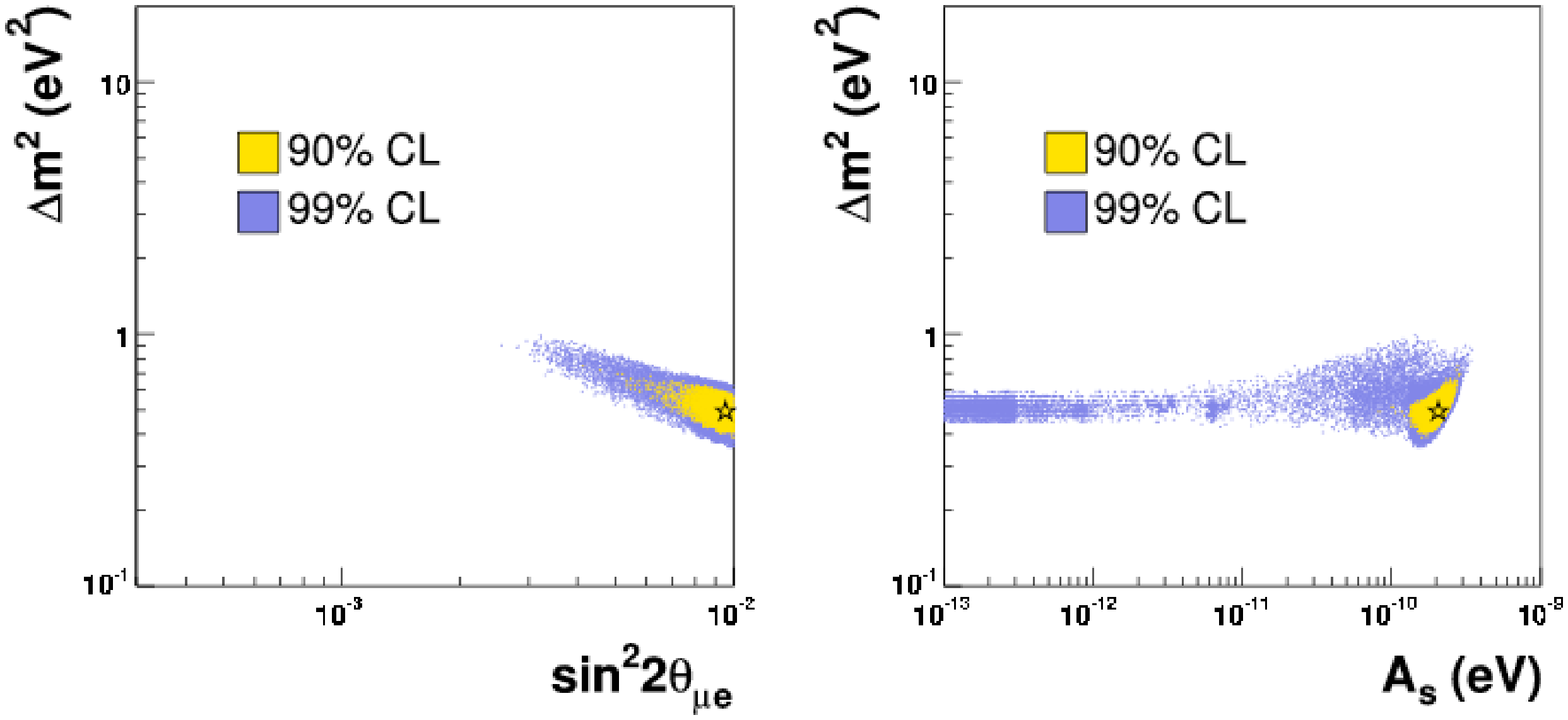}
\end{center}
\caption{\label{allowedmat}The 90\% and 99\% confidence level (CL) allowed vaccum oscillation parameters and $A_s$, obtained from a joint (3+1) with matter effect fit to MB$\nu$, MB$\bar{\nu}$ and LSND data sets.}
\end{figure}

The allowed parameters at 90\% and 99\% CL are shown in Fig.~\ref{allowedmat}. The best-fit parameters, indicated on the figure in black stars, correspond to
\begin{eqnarray}
\label{bf}
\sin^22\theta_{\mu e}=0.010~, \nonumber \\
\Delta m^2=0.47\mathrm{\ eV}^2~, \nonumber \\
A_s = 0.2\times10^{-10}\mathrm{\ eV}~.
\end{eqnarray}

Note that the $\Delta m^2$ and $\sin^22\theta_{\mu e}$ values quoted above correspond to the vacuum oscillation parameters, and are consistent with a fourth, mostly sterile neutrino mass eigenstate with $\Delta m^2_{41}\gg\Delta m^2_{32},\Delta m^2_{21}$, by construction. The effective best-fit parameters for each of the three data sets considered in the fit are summarized in Tab.~\ref{tab}.

\begin{table}
\begin{center}
\begin{tabular}{l|c|c}
\hline
Data Set & $\Delta m^2_M$ (eV$^2$) & $\sin^22\theta_{\mu e}^M$ \\ \hline\hline
MB$\nu$ & 0.75 & 0.0016 \\
MB$\bar{\nu}$ & 0.22 & 0.08 \\
LSND & 0.45 & 0.012 \\ \hline
\end{tabular}
\caption{\label{tab}Effective best-fit mixing parameters for the data sets considered in the fit. The parameters have been calculated using $E_{MB\nu}=600$~MeV, $E_{MB\bar{\nu}}=700$~MeV and $E_{LSND}=40$~MeV. $A_s$ corresponds to the best-fit value of 2.0$\times$10$^{-10}$~eV.}
\end{center}
\end{table}

\section{\label{sec:fivec}QUALITATIVE DISCUSSION OF RESULTS}

The (3+1) fit results are straightforward to interpret. In the case of no matter effects, the single frequency ($\Delta m^2$) involved in a two-neutrino oscillation approximation we have employed in these fits is unable to reconcile the MiniBooNE neutrino mode excess with the MiniBooNE and LSND antineutrino excesses, as they show up at different $L/E$ (see Fig.~\ref{LE}). 

Introducing a matter effect potential which flips sign when going from neutrino to antineutrino oscillations, allows for modifying the location ($\Delta m^2_M$) and amplitude ($\sin^22\theta^M_{\mu e}$) of the oscillation maximum for neutrinos and antineutrinos independently, as well as as a function of $E$ (and $L$). Thus, oscillation probability measurements performed at the same $L$ and $E$ can yield different values for the amplitude and location of the oscillation maximum  and corresponding observed excess depending on whether they are performed using neutrinos or using antineutrinos. Furthermore, measurements at different $E$ can yield excesses which point to different  amplitude and oscillation maximum location even if they are extracted with the same polarity beam. 

\begin{figure}[h]
\begin{center}
\includegraphics[width=4in, trim=0 0 0 0]{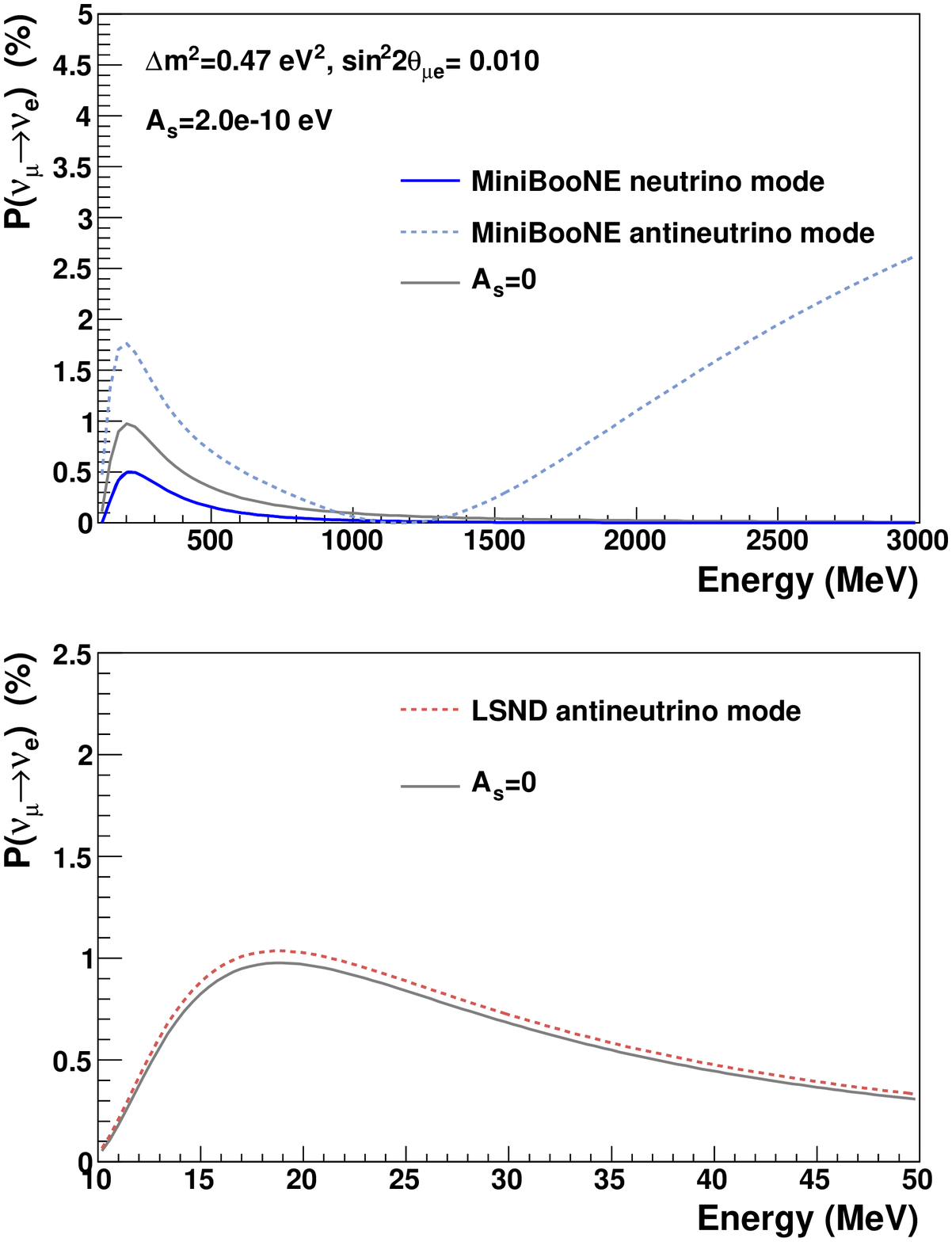}
\end{center}
\caption{\label{appprobEsbl}{Effective appearance oscillation probabilities for MiniBooNE (top) and LSND (bottom) as a function of neutrino energy, calculated using the best-fit values in Eq.~\ref{bf}. The gray lines correspond to the same oscillation parameters but with $A_s$ set to zero.}}
\end{figure}

Figure \ref{appprobEsbl} is instructive in understanding how this matter effect leads to the distributions shown in Fig.~\ref{distributions}. The top panel shows the expected $\nu_{\mu}\rightarrow\nu_e$ and $\bar{\nu}_{\mu}\rightarrow\bar{\nu}_e$ oscillation probabilities at MiniBooNE, as a function of neutrino energy. The gray line corresponds to the best-fit oscillation parameters but with $A_s=0$, while the blue solid and dashed lines correspond to the best-fit parameters and $A_s=2.0\times10^{-10}$~eV. Both oscillation probabilities suggest an excess at low energy; however, a non-zero $A_s$ value allows for oscillations at higher energy at the $\sim$1\% level in the case of antineutrinos, and lack of oscillations at higher energy in the case of neutrinos, consistend with MiniBooNE observations. The bottom panel shows the expected $\bar{\nu}_{\mu}\rightarrow\bar{\nu}_e$ oscillation probability at LSND. Here, a $\sim$1\% oscillation probability is preserved with or without a non-zero $A_s$, due to the much lower energy range. 

\begin{figure}[h]
\begin{center}
\includegraphics[width=4in, trim=0 0 0 0]{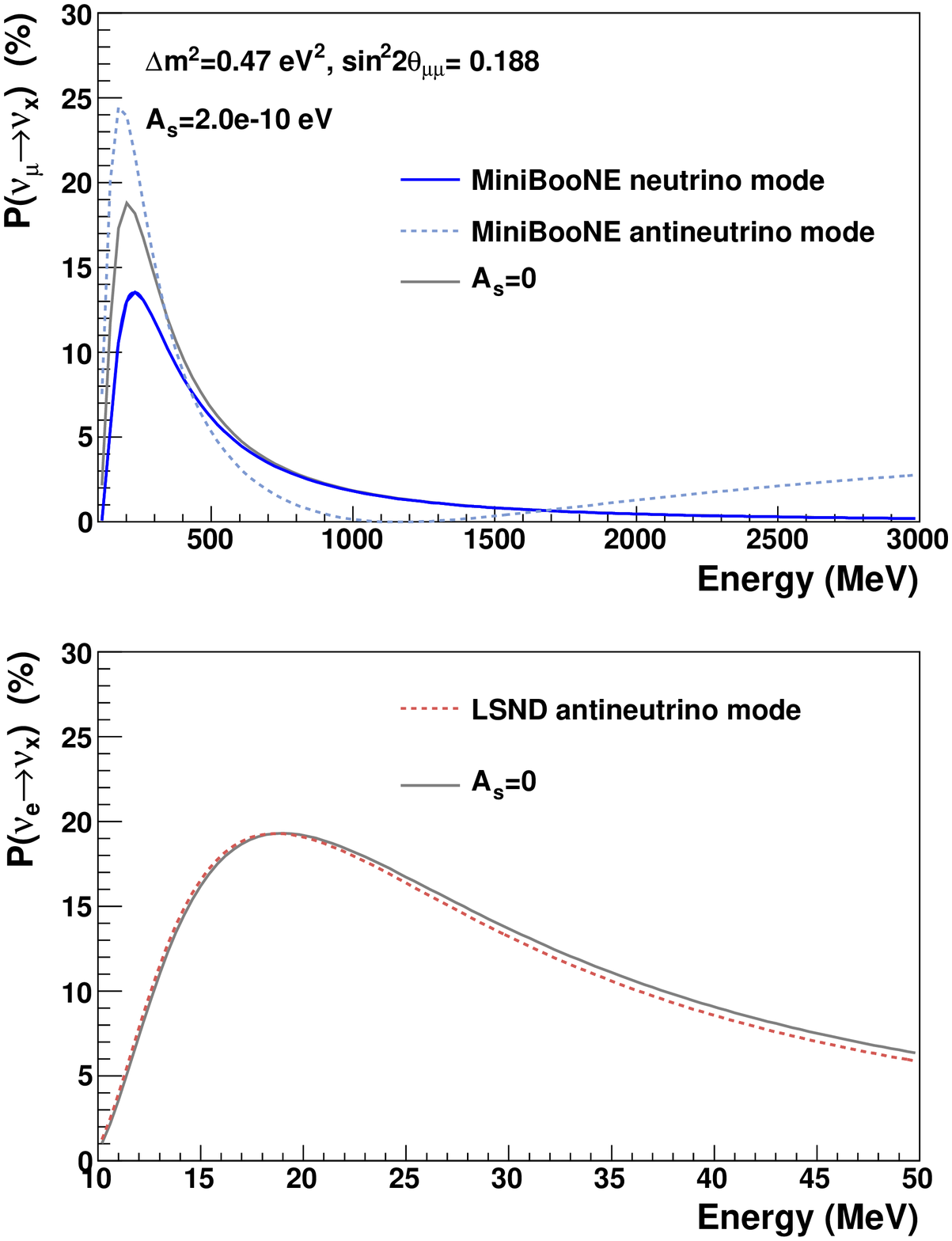}
\end{center}
\caption{\label{disprobEsbl}{Effective disappearance oscillation probabilities for MiniBooNE (top) and LSND (bottom) as a function of neutrino energy, calculated using the best-fit values in Eq.~\ref{bf}. The gray lines correspond to the same oscillation parameters but with $A_s$ set to zero.}}
\end{figure}

Also interesting to explore are the $\stackrel{\tiny{(-)}}{\nu}_{\mu}$ and $\stackrel{\tiny{(-)}}{\nu}_e$ disappearance probabilities expected in this oscillation framework, 
\begin{equation}
P(\stackrel{\tiny{(-)}}{\nu}_{\alpha}\rightarrow\stackrel{\tiny{(-)}}{\nu}_{\not{\alpha}})=\sin^22\theta^M_{\alpha\alpha}\sin^2(1.27\Delta m^2_M L/E)~,
\end{equation}
which can be calculated using 
\begin{equation}
\sin^22\theta_{\alpha\alpha}^M=4|U^M_{\alpha4}|^2(1-|U_{\alpha4}^M|^2)~,
\end{equation}
and the definitions in Eqs.~\ref{effu4} and \ref{effum4}. In the case of MiniBooNE, the expected $\nu_{\mu}$ and $\bar{\nu}_{\mu}$ disappearance probabilities corresponding to the best-fit parameters of Eq.~\ref{bf} are shown in Fig.~\ref{disprobEsbl} (top). In both cases (neutrino and antineutrino), one expects disappearance probabilities on the order of a few percent, on average. The oscillation probability peaks at lower energies, where MiniBooNE $\nu_{\mu}$ and $\bar{\nu}_{\mu}$ disappearance searches become dominated by flux and cross section uncertainties \cite{AguilarArevalo:2009yj,Mahn:2011ea}. The upcoming joint MiniBooNE/SciBooNE $\bar{\nu}_{\mu}$ disappearance search, however, may have some sensitivity to this effect. Similarly, the bottom panel of Fig.~\ref{disprobEsbl} shows the $\nu_e$ disappearance probability expected at LSND, corresponding to the best-fit parameters of Eq.~\ref{bf}. The expected oscillation probability for $E\sim30-40$~MeV is approximately 10\%. The KARMEN and LSND $\nu_e$-Carbon charged-current cross section measurements which are available in this energy range do not have sufficient sensitivity to address this \cite{Conrad:2011ce}.

\begin{figure}[h]
\begin{center}
\includegraphics[width=4in, trim=0 0 0 0]{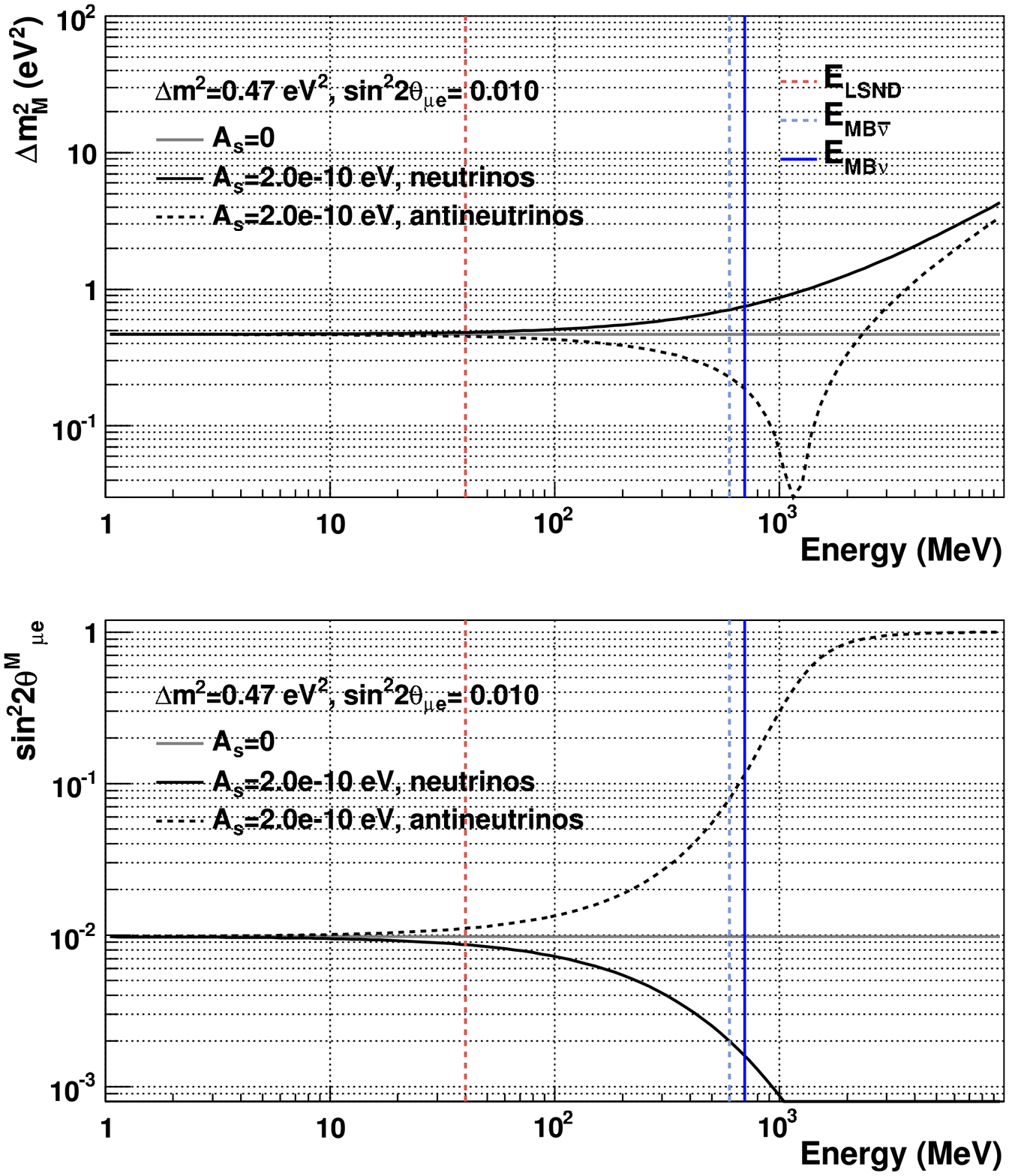}
\end{center}
\caption{\label{effmixvsE}{The effective $\Delta m^2_M$ and $\sin^22\theta_{\mu e}^M$ mixing parameters corresponding to the best-fit values in Eq.~\ref{bf}, each as a function of neutrino energy. The y-axis parameter at the point of intersection of any (dashed or solid) vertical line with the corresponding (dashed or solid, respectively) black line corresponds to the effective $\Delta m^2_M$ parameter seen by a particular data set, as listed in Tab.~\ref{tab}. For example, in the top figure, the $E_{LSND}$ vertical line, in dashed red, intersects with the dashed black line at $\Delta m^2_M=0.45$~eV$^2$, and so the effective mass-squared difference seen by a 40~MeV antineutrino in the LSND beam corresponds to 0.45 eV$^2$.}}
\end{figure}

Figure \ref{effmixvsE} is instructive in understanding how the effective oscillation parameters vary with neutrino energy. The top panel of Fig.~\ref{effmixvsE} shows the effective $\Delta m^2_M$ oscillation parameter in matter as a function of neutrino energy, for the best-fit oscillation and $A_s$ values. The solid black line indicates the $\Delta m^2_M$ dependence on $E$ in the case of a neutrino beam; the dashed line shows the antineutrino dependence. The mean neutrino energy for each data set is also shown by a vertical line. Similarly, the bottom panel of Fig.~\ref{effmixvsE} shows the effective $\sin^22\theta^M_{\mu e}$ oscillation parameter in matter, and how it varies with neutrino energy, assuming the best-fit parameters obtained in Sec.~\ref{sec:fiveb}. It is evident from this figure that, in the case of neutrino oscillation experiments performed in the few MeV to few GeV range, one should not expect large appearance amplitudes, since the effective mixing amplitude is always less than that expected from the underlying vacuum parameter, $\sin^22\theta_{\mu e}=0.010$. On the other hand, one expects to observe resonance-like effects with antineutrino experiments performed in the $>1$~GeV range.

As a final discussion point, we remark on the size of $A_s$. While the form of the matter potential we have assumed is the same as that of the standard matter effect in three-neutrino oscillations ($A_{SM}$=constant), the resulting best-fit value suggested by the MiniBooNE and LSND data sets is roughly 500 times larger than that of standard matter effect. The result is consistent with the analytical picture presented in \cite{Kamo:2002sj}, where the authors state that in order to obtain observable, non-negligible matter effects in a model with a single sterile neutrino, one needs a matter effect potential of $A_s\simeq10^{-10}$~eV. The large value of $A_s$ may be quite difficult to explain in simple extensions to the theory. 
However, the good agreement of all three signatures within this three-parameter model motivates the development of theoretical interpretations which would lead to such a phenomenological effect.

\section{\label{sec:six}EXPERIMENTAL IMPLICATIONS FOR SOLAR AND ATMOSPHERIC OSCILLATIONS}


Even though the vacuum oscillation parameters allowed in this model are constructed so as not to interfere with the atmospheric and solar oscillation scales, the $E$ and $V_s$ dependence can drive the effective $\Delta m^2_M$ and $\sin^22\theta^M_{\alpha\beta}$ parameters to the solar and atmospheric parameter regions. Therefore, one must ask whether constraints arise from solar/reactor and atmospheric/long-baseline accelerator experiments.

Reactor experiments pose little to no constraints to this scenario, as the low energy assures that the effective oscillation parameters deviate very little from the vacuum parameters (which, according to the best fit, correspond to $\Delta m^2=0.47$~eV$^2$ and $\sin^22\theta_{ee}=0.19$). In fact, the reactor anomaly recently identified in short-baseline reactor experiments is consistent with small-amplitude oscillations due to a heavy ($\ge$1 eV$^2$) mostly-sterile mass eigenstate \cite{reactor}, and could reasonably be accommodated within the matter effect scenario. Joint fits including the reactor data are necessary for more quantitative tests.

It is likely that such matter effects affect neutrino propagation in the sun. However, this is probably dependent on the nature of $V_s$. Such effects should be investigated for particular $V_s$ underlying scenarios.

\begin{figure}
\begin{center}
\includegraphics[width=3in, angle=-90, trim=0 0 0 0]{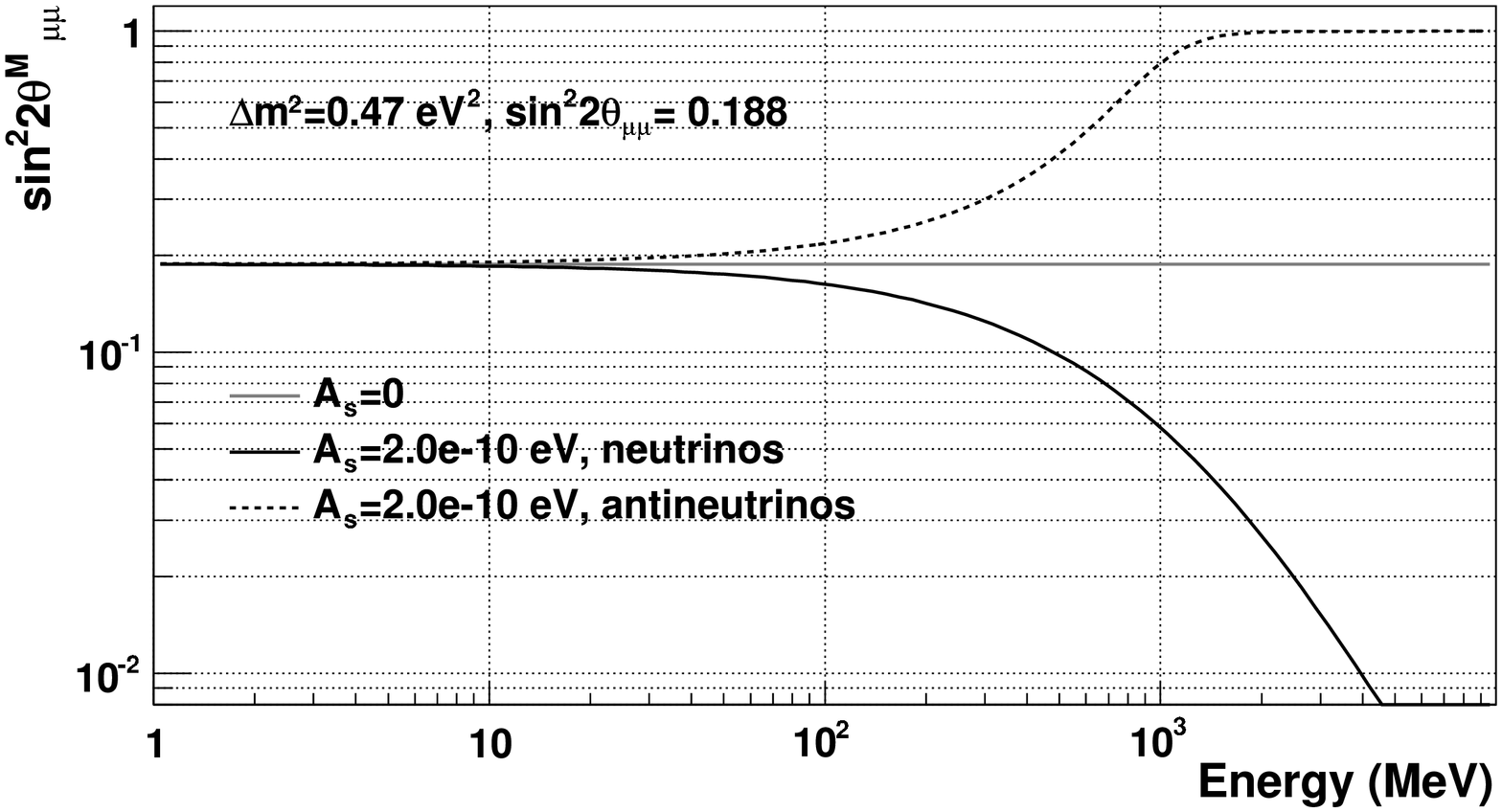}
\end{center}
\caption{\label{effs2thmm}Dependence of the effective disappearance amplitude $\sin^22\theta_{\mu\mu}^M$, corresponding to the best-fit values in Eq.~\ref{bf}, on neutrino energy.}
\end{figure}

In the case of atmospheric/long-baseline accelerator oscillation experiments, we use the MINOS experiment as an example to investigate the effects of the best-fit model in Sec.~\ref{sec:fiveb}. MINOS employs a near/far detector setup at $\sim0.75/735$ km from a neutrino source, in order to look for $\nu_{\mu}$ and $\bar{\nu}_{\mu}$ disappearance driven by the atmospheric $\Delta m^2$ scale, $\Delta m^2_{32}$. The MINOS neutrino energy spectrum spans the 1-10~GeV energy range, with a peak neutrino energy at 3 GeV. Because the $\Delta m^2_M$ deviation from the vacuum oscillation parameter becomes larger, overall, with increasing energy, and MINOS sits at a much higher energy than MiniBooNE and LSND, one might expect noticeable effects. From Fig.~\ref{effmixvsE}, one sees that for $E\sim3$ GeV, the effective $\Delta m^2_M$ for neutrinos is similar to or larger than the vacuum $\Delta m^2$, and therefore one might expect any possible oscillations to average out to half the disappearance amplitude, $\sin^22\theta_{\mu\mu}^M/2$, for both the near and far MINOS detectors. Here, $\sin^22\theta_{\mu\mu}^M$ corresponds to
\begin{equation}
\sin^22\theta_{\mu\mu}^M=4|U^M_{\mu4}|^2(1-|U_{\mu4}^M|^2)~,
\end{equation}
which is the standard $\nu_{\mu}$ disappearance probability amplitude for (3+1) sterile neutrino oscillations, except with $U_{\mu}\rightarrow U_{\mu4}^M$, from Eq.~\ref{effum4}. Figure~\ref{effs2thmm} shows the $\sin^22\theta_{\mu\mu}^M$ dependence on $E$ for the best-fit parameters. One sees that $\sin^22\theta_{\mu\mu}^M$ is always equal to or smaller than the vacuum oscillation parameter $\sin^22\theta_{\mu\mu}=0.19$ for any neutrino oscillation experiment. Therefore, the MINOS neutrino mode $\nu_{\mu}$ data set is insensitive to this model, since systematic uncertainties on the absolute event rate prediction at the near detector are much larger than this level.

\begin{figure}[h]
\begin{center}
\includegraphics[width=6in, trim=0 0 0 0]{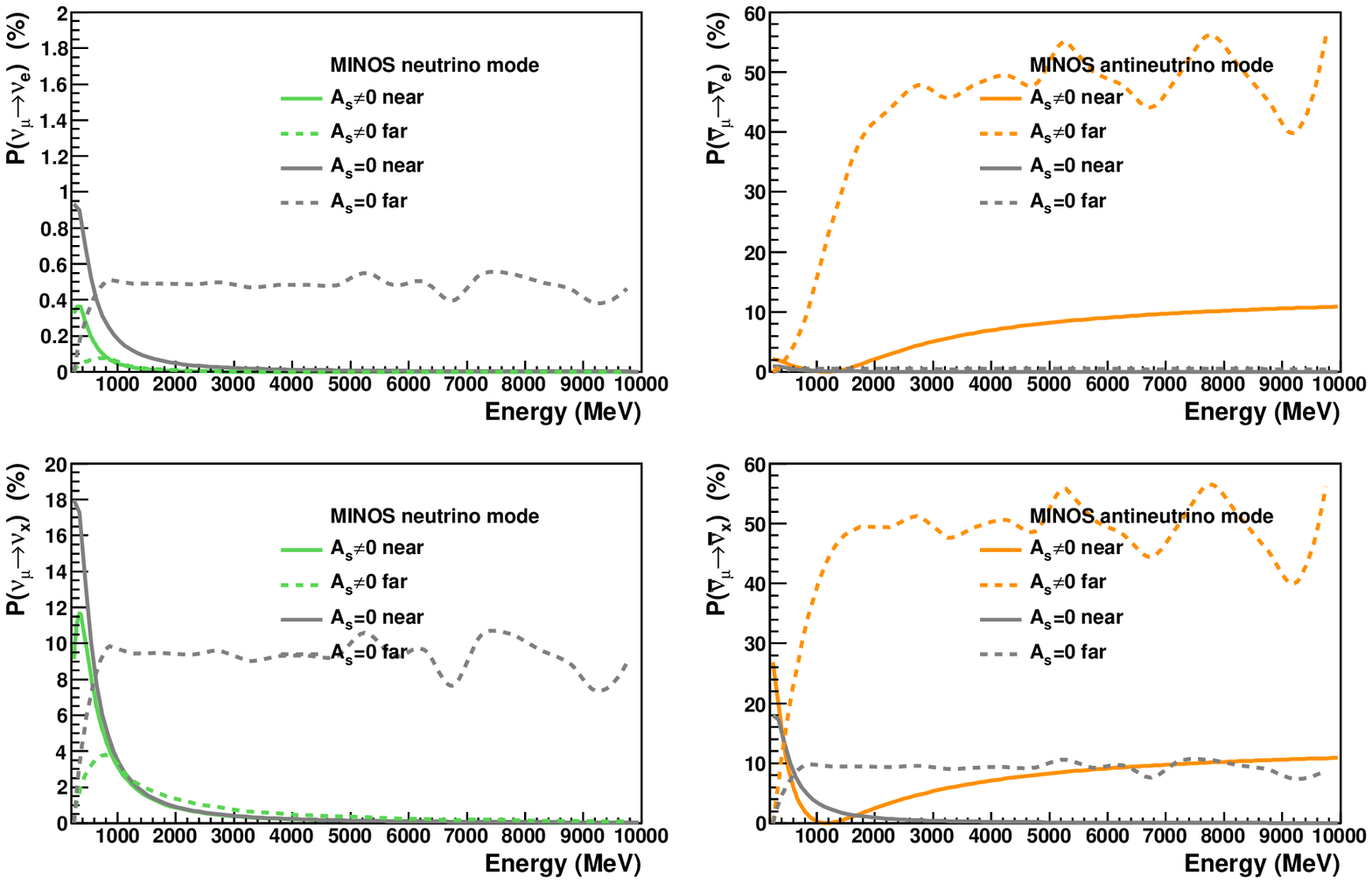}
\end{center}
\caption{\label{appprobEminos}{Effective $\stackrel{\tiny{(-)}}{\nu}_{\mu}\rightarrow\stackrel{\tiny{(-)}}{\nu}_e$ appearance (top) and $\stackrel{\tiny{(-)}}{\nu}_{\mu}$ disappearance (bottom) oscillation probabilities for MINOS as a function of neutrino energy, calculated using the best-fit values in Eq.~\ref{bf}. The left figures show neutrino mode oscillation probabilities. The right figures show antineutrino mode oscillation probabilities. The gray lines correspond to the same oscillation parameters but with $A_s$ set to zero.}}
\end{figure}

In the case of antineutrino oscillations, however, the $\Delta m^2_M$ value decreases with $E$ and becomes negative beyond $E_0=\Delta m^2/(2A_s)\simeq1180$~MeV. Note, however, that $\sin^2(1.27\Delta m^2_M L/E)$ is insensitive to the sign of $\Delta m_M^2$. Beyond this point, therefore, the effective $|\Delta m^2_M|$ starts increasing linearly with $E$. This is illustrated in Fig.~\ref{effmixvsE}. Given the best-fit oscillation parameters, we see that $|\Delta m^2_M|$ becomes close to the atmospheric $\Delta m^2_{32}$ for a narrow range of energies around $E_\sim1200$~MeV. This range is on the very low edge of the MINOS energy range, and while the $\sin^22\theta_{\mu\mu}^M$ value for antineutrinos at this energy is predicted to be nearly maximal, without an actual fit it is difficult to make quantitative statements as to the level of constraint provided by MINOS antineutrino data, since the $\Delta m^2_M$ and $\sin^22\theta_{\mu\mu}^M$ vary rapidly in the MINOS antineutrino energy range. On the other hand, it is reasonable to expect differences in the neutrino and antineutrino oscillation parameters obtained from MINOS $\nu_{\mu}$ and $\bar{\nu}_{\mu}$ disappearance searches, respectively, such as those presented in \cite{Adamson:2011ch,Adamson:2011fa}. The bottom panels of Fig.~\ref{appprobEminos} show the observable $\stackrel{\tiny{(-)}}{\nu}_{\mu}$ disappearance probabilities expected at MINOS in neutrino and antineutrino mode at both the far and near detectors. 

The top panels of Fig.~\ref{appprobEminos} show the corresponding observable $\stackrel{\tiny{(-)}}{\nu}_{\mu}\rightarrow\stackrel{\tiny{(-)}}{\nu}_{e}$ appearance probabilities expected at MINOS near and far detectors \footnote{In the case of antineutrino oscillations, we obtain rapid oscillations in the far detectors, and so the corresponding oscillation probabilities shown in these plots have been averaged out.}. In neutrino mode, one expects a $<1$\%  appearance oscillation probability across the 1-10~GeV energy range at the far detector, which is far beyond MINOS' sensitivity, but in antineutrino mode, on expects large appearance effects which may be measurable in a far to near comparison, given enough statistics.

\section{\label{sec:seven}CONCLUSIONS}

We have examined MiniBooNE and LSND results in oscillation fits to a model with a single, mostly-sterile neutrino mass eigenstate at a $\Delta m^2\sim1$ eV$^2$, with and without the presence of an effective matter-like potential of the form
\begin{equation}
V_s=\pm A_s~,
\end{equation}
experienced only by sterile neutrino/antineutrino states. We find that the compatibility among LSND and MiniBooNE neutrino and antineutrino data sets increases significantly in the presence of a non-zero $A_s$, with the best-fit parameters corresponding to the vacuum mixing parameters $\sin^22\theta_{\mu e}=0.010$ and $\Delta m^2=0.47$~eV$^2$, and $A_s=2.0\times10^{-10}$~eV. The best-fit parameters are consistent with reactor long-baseline, atmospheric, and accelerator long-baseline (neutrino) results, and can reasonably accommodate the recent reactor short-baseline anomalous result. Implications for the MINOS antineutrino data set have also been considered; the best-fit model predicts some observable effects in the MINOS antineutrino samples, and therefore MINOS' sensitivity to those effects should be explored further. We invite phenomenological interpretations of this model.


\begin{acknowledgments}
\noindent We thank Andre de Gouvea for valuable discussions. We also thank the National Science Foundation for their support.
\end{acknowledgments}

%


\end{document}